\documentclass[12pt]{article}
\usepackage{amsmath,epsf,amssymb,latexsym,cite,xy}
\xyoption{matrix}\xyoption{arrow}\xyoption{frame}

\newtheorem{theorem}{Theorem}

\newif\iffigs\figstrue

\DeclareFontFamily{U}{rsf}{}
\DeclareFontShape{U}{rsf}{m}{n}{
  <5> <6> rsfs5 <7> <8> <9> rsfs7 <10-> rsfs10}{}
\DeclareMathAlphabet\Scr{U}{rsf}{m}{n}

\def\pplogo{\vbox{\kern-\headheight\kern -43pt
\halign{##&##\hfil\cr&{
\ppnumber}\cr\rule{0pt}{2.5ex}&\ppdate\cr}
}}
\makeatletter
\def\ps@firstpage{\ps@empty \def\@oddhead{\hss\pplogo}%
  \let\@evenhead\@oddhead 
}
\def\maketitle{\par
 \begingroup
 \def\thefootnote{\fnsymbol{footnote}}
 \def\@makefnmark{\hbox{$^{\@thefnmark}$\hss}}
 \if@twocolumn
 \twocolumn[\@maketitle]
 \else \newpage
 \global\@topnum\z@ \@maketitle \fi\thispagestyle{firstpage}\@thanks
 \endgroup
 \setcounter{footnote}{0}
 \let\maketitle\relax
 \let\@maketitle\relax
 \gdef\@thanks{}\gdef\@author{}\gdef\@title{}\let\thanks\relax}
\makeatother

\def\grade{{\varphi}}
\def\O{\Scr{O}}
\def\C{{\mathbb C}}
\def\P{{\mathbb P}}

\def\R{{\mathbb R}}
\def\Z{{\mathbb Z}}

\def\Im{\operatorname{Im}}
\def\Hom{\operatorname{Hom}}

\def\Ext{\operatorname{Ext}}

\def\rank{\operatorname{rank}}

\def\Cone{\operatorname{Cone}}

\def\ch{\operatorname{\mathit{ch}}}
\def\td{\operatorname{\mathit{td}}}

\def\delbar{{\bar\partial}}

\def\CY{Calabi--Yau}

\def\cR{{\Scr R}}
\def\cM{{\Scr M}}
\def\cN{{\Scr N}}
\def\CN{\cN}
\def\cA{{\Scr A}}

\def\cI{{\Scr I}}

\def\cP{{\Scr P}}
\def\cT{{\Scr T}}

\def\cE{{\Scr E}}
\def\cF{{\Scr F}}
\def\cX{{\Scr X}}
\def\cG{{\Scr G}}
\def\DC{\mathbf{D}}

\def\ff#1#2{{\textstyle\frac{#1}{#2}}}
\def\half{\frac{1}{2}}

\def\Lotimes{\mathrel{\mathop\otimes^{\mathbf{L}}}}

\begin{document}
\setcounter{page}0
\def\ppnumber{\vbox{\baselineskip14pt
\hbox{DUKE-CGTP-01-13}
\hbox{RUNHETC-2001-29}
\hbox{hep-th/0110071}}}
\def\ppdate{October 2001} \date{}

\title{\LARGE D-Brane Stability and Monodromy\\[10mm]}
\author{
Paul S.~Aspinwall\\[2mm]
\normalsize Center for Geometry and Theoretical Physics \\
\normalsize Box 90318 \\
\normalsize Duke University \\
\normalsize Durham, NC 27708-0318\\[8mm]
Michael R.~Douglas\\[2mm]
\normalsize Department of Physics and Astronomy \\
\normalsize Rutgers University \\
\normalsize Piscataway, NJ 08855-0849 \\
\normalsize {\it and} \\
\normalsize Institut des Hautes \'Etudes Scientifiques \\
\normalsize Bures-sur-Yvette, 91440 France
}

{\hfuzz=10cm\maketitle}

\def\Large{\large}
\def\LARGE{\large\bf}


\begin{abstract}
We review the idea of $\Pi$-stability for B-type D-branes on
a \CY\ manifold. It is shown that the octahedral axiom from the
theory of derived categories is an essential ingredient in the study
of stability. Various examples in the context of the quintic \CY\
threefold are studied and we plot the lines of marginal stability in
several cases.  We derive the conjecture of
Kontsevich, Horja and Morrison for the derived category version of
monodromy around a ``conifold'' point.
Finally, we propose an application of these ideas
to the study of supersymmetry breaking.
\end{abstract}

\vfil\break


\section{Introduction}    \label{s:intro}

How exactly are we to consider a D-brane on a \CY\ space $X$? The basic
picture would be some cycle $C\subset X$ perhaps equipped with some kind
of vector bundle $E\to C$. There has to be more to D-branes than this
however. It has been known since the early days of mirror symmetry,
from works such as Candelas et al \cite{CDGP:}, that the general notion of a
cycle $C\subset X$ is far from clear and perhaps not even
well-defined.

Suppose one begins with an even-dimensional cycle lying in some
homology class. The simplest example would be a point whose homology
class lies in $H_0$. Upon
``parallel transport'' around a loop in the moduli space of the
complexified K\"ahler form, $B+iJ$, this homology class is generally
transformed into some other element of $H^{\textrm{even}}$. Thus the
point appears to have transformed magically into a general collection
of even-dimensional cycles.

Based in the ideas of Mukai \cite{Muk:FM,Muk:FM2}, Kontsevich
\cite{Kon:mir} realized that this monodromy appeared fairly naturally in
the more abstract setting of the derived category of coherent
sheaves. A more direct analysis of D-branes in terms of the derived
category was then pioneered in \cite{Doug:DC} together with further
analysis in \cite{AL:DC,Dia:DC,Laz:DC}.

One of the purposes of this paper is to show explicitly what happens
physically to the D-branes as we undergo such monodromy. We focus purely
on even-dimensional (B-type) D-branes. In order to do this we have to
first study the notion of stability of a D-brane which is itself a
very interesting subject. 

The notion of stability we use is 
the $\Pi$-stability of \cite{DFR:stab}. 
As generalized in \cite{Doug:DC},
this is a criterion determining which elements in the
derived category actually correspond to physical BPS D-branes, on any
Calabi-Yau with any moduli.  The generality of this proposal comes
with a price, and many of the issues which come up in practice remain
to be resolved.  We discuss some of these, and illustrate them with
several examples of D-branes on the quintic \CY\ threefold.  Some of
the points we raise were also discussed in \cite{Doug:S01}.  Other
recent work on D-branes on Calabi-Yau's includes
\cite{DGR:Dquin,HKLM:LS,dFS:MK,LMW:McK,GJ:CohD}.

It is perhaps worth emphasizing here that the subjects of the derived
category and $\Pi$-stability are probably central to an understanding
of many aspects of string theory. While we focus here purely on
D-branes, it was observed in \cite{AD:tang} that the derived
category, and presumably $\Pi$-stability, are fairly unavoidable in
the heterotic string. (See \cite{KS:toricF,Sharpe:subK}, for example,
for some comments on stability in the classical sense.)
Indeed this kind of approach may well be
required for the complete study of any Yang--Mills (or at least
Hermitian--Yang--Mills) theory in a stringy context.

In section~\ref{s:pistab} we review the basic ideas of the origin of
the derived category in the context of D-branes and the appearance of
$\Pi$-stability. In order to obtained a well-defined notion of
stability we show that the ``octahedral axiom'' of triangulated
categories in required.

In section~\ref{s:eg} we clarify the analysis of
section~\ref{s:pistab} by applying it to the quintic \CY\
threefold. This section contains several plots showing the
lines of marginal stability of several D-branes.

In section~\ref{s:mon} we show how monodromy manifests itself
physically as a result of stability considerations. 

Finally, in section~\ref{s:susy} we discuss an application 
to the study of supersymmetry breaking.


\section{$\Pi$-Stability}   \label{s:pistab}

In this section we review and elaborate on the notions of
$\Pi$-stability for objects in the derived category of coherent
sheaves. The course of logical steps required to understand
$\Pi$-stability is somewhat complicated so we will try to spell out each
step of the argument.


\subsection{Topological D-Branes} \label{ss:top}

Let us begin with the theory of topological B-branes as discussed in
\cite{Doug:DC,AL:DC}.  The definition of the topological B model is
at first sight independent of K\"ahler moduli, and at first sight 
topological B branes will
be just the boundary conditions discussed by Witten  \cite{W:CS},
associated with holomorphic cycles carrying holomorphic bundles.

However, this is not all of the topological boundary conditions.  This
point can be motivated in many ways.  For example, the typical \CY\
stringy K\"ahler moduli space as determined by mirror symmetry has
several large volume limits producing spaces of distinct topology; this
leads to several distinct sets of holomorphic bundles, while the
complete set of boundary conditions must be the same in all such
limits.

There is a natural mathematical construction which, starting from any of
these sets (actually categories) of boundary conditions, leads to a universal
result: to pass from the original category to the derived category.
Implicit in Kontsevich's homological mirror symmetry proposal \cite{Kon:mir}
is the claim that topological branes correspond precisely to objects
in the derived category.  The more recent proposals of \cite{Doug:DC,AL:DC}
explain physically how to construct such boundary conditions, and why this
generalization is necessary.

Initially the boundaries in Witten's B-model are taken to be a set of
finite-dimensional vector bundles. 
We then further distinguish boundary conditions by
associating a number $n\in\Z$
known as the ``grade'', to each possible boundary.
Let us denote a bundle associated with a boundary
labeled by $n$ as $E^n$. 

A topological open string stretching from $E^m$ to $F^n$ then lives in
the Hilbert subspace  
$H^p(X,(E^m)^\vee\otimes F^n)$, where $p$ is the intrinsic ``bulk''
ghost number of the open string.
At this point we have used the one-to-one 
relation between states in topological
string theory, Ramond ground states in the corresponding physical
string (and thus massless fermions in space-time), and to the
(Dolbeault) cohomology of the $\delbar$ operator coupled to the bundle
$(E^m)^\vee\otimes F^n$.

The total ghost
number in the topological field theory is then redefined as follows:
an open string between branes with abstract grades $m$ and $n$, i.e.
in $H^p(X,(E^m)^\vee\otimes F^n)$, is
{\em defined\/} to have ghost number $p-m+n$.

The wedge product gives us an associative product on cohomology:
\begin{equation}
H^p(X,(E^m)^\vee\otimes F^n) \times H^q(X,(F^n)^\vee\otimes G^p)
\to H^{p+q}(X,(E^m)^\vee\otimes G^p).
\end{equation}
In topological field theory terms this is the
operator product.  
At this point we switch to the language
of algebraic geometry.\footnote{It would be interesting to know if any of these
considerations can be generalized for differential geometry and less or
no supersymmetry.}
We replace the notion of holomorphic vector bundle with the associated
coherent sheaf. These sheaves, to be thought of as D-branes, together
with their morphisms, to be thought of as open strings, form the
structure of a category.

This category is in fact an abelian category: this means that every map between
sheaves has a kernel and cokernel.
Gaining this property is one reason to generalize from bundles to sheaves; 
it is only true in
this larger context.  Physically, the first point at which sheaves enter
is in considering the zero size instanton, as explained in \cite{HM:alg2}.
Thus, the vector bundle $E^n$ is generalized to a
sheaf $\cE^n$, and the linear vector space $H^p(X,(E^m)^\vee\otimes E^n)$
becomes the linear vector space $\Ext^p(\cE^m, \cE^n)$.

The next step is to allow objects $E$ which are
direct sums of objects $E^n$ with different abstract grades $n$, and to
allow a BRST operator which is a general map $Q:E\to E$ of grade $1$
satisfying $Q^2=0$.  In other words, each B-brane can be written
as a complex of finite length:
\begin{equation}
\xymatrix@1{
\ldots\ar[r]&\cE^{-1}\ar[r]^{d_{-1}}&\underline{\cE^0}\ar[r]^{d_0}
&\cE^1\ar[r]^{d_1}&\ldots ,
} \label{eq:Ecplx}
\end{equation}
where the boundary maps $d^n$ represent the various components of the
BRST operator $Q$.  Such a generalization respects
all of the axioms of topological field theory; it can be further argued
that this generalization is in fact the general ghost number one 
deformation of the topological field theory \cite{AL:DC}.\footnote{
See \cite{Dia:DC} for a
further generalization which uses non-holomorphic forms,
$\delbar \omega\ne 0$.  At present there is no evidence that this
further generalization is required to describe BPS states.}
Some notations from \cite{AL:DC} are used as follows: an
underline in a complex represents the
``zeroth'' position. Also, if $\cF$ is a coherent sheaf then
$\underline\cF$ will represent a complex with the only nonzero entry
being $\cF$ at the zeroth position. We also use the conventional
notation that $\cE^\bullet[q]$ represents the complex $\cE^\bullet$
shifted $q$ places to the {\em left}.

We will denote the complex (\ref{eq:Ecplx}) as $\cE^\bullet$.
Note that the original notion of ``grade'' of a sheaf has now become
``position'' in a complex.

As argued in \cite{Doug:DC}, many complexes correspond to the same
topological brane, distinguished by adding brane-antibrane pairs or
other combinations of branes which can annihilate to the vacuum.  To
factor out this redundancy, two complexes are taken to represent the
same D-brane if and only if they are related by a {\em
quasi-isomorphism}. This equivalence was explained in a slightly
different way in \cite{AL:DC}. The final result is that topological
B-branes are (up to isomorphism) in one to one correspondence with
objects in $\DC(X)$ --- the bounded derived category of coherent
sheaves on $X$.


\subsection{Cones and Triangles}  \label{ss:tri}

Further marginal deformations of the topological field theory of
section \ref{ss:top} yields nothing new in the sense that the
resulting D-branes remain objects in $\DC(X)$.
Given a pair of complexes $\cE^\bullet$ and $\cF^\bullet$ together
with a chain map $f:\cE^\bullet\to\cF^\bullet$, the ``mapping cone''
$\Cone(f:\cE^\bullet\to\cF^\bullet)$ is defined as an object in
$\DC(X)$ given by
\begin{equation}
\xymatrix@1@C=15mm{
\ar[r]&\underline{\cE^1\oplus\cF^0}\ar[r]^{\left(\begin{smallmatrix}d_E&0\\
f&d_F \end{smallmatrix}\right)}&
\cE^2\oplus\cF^1\ar[r]^-{\left(\begin{smallmatrix}d_E&0\\
f&d_F \end{smallmatrix}\right)}&
\cE^3\oplus\cF^2\ar[r]&\ldots
}  \label{eq:cone}
\end{equation}
It was argued in \cite{AL:DC} that this is a general feature of
topological D-branes. Namely two D-branes $\cE^\bullet,\cF^\bullet$ in
$\DC(X)$ are marginally deformed by a map $f$ into
$\Cone(f:\cE^\bullet\to\cF^\bullet)$. Thus the mapping cone represents
the ``sum'' of two branes.

In the simplest case of the mapping cone we may think of $\cE^\bullet$
as a the complex given by the single sheaf $\cE$ at position zero and 
$\cF^\bullet$
as a the complex given by the single sheaf $\cF$ at position zero. A
map $f:\cE\to\cF$ then yields the mapping cone as a complex:
\begin{equation}
\xymatrix@1{
0\ar[r]&\cE\ar[r]^{f}&\underline{\cF}\ar[r]&0.
}
\end{equation}

Reversing this construction one may view the complex
\begin{equation}
\xymatrix@1{
\ldots\ar[r]&\cE^{-1}\ar[r]^{d_{-1}}&\underline{\cE^0}\ar[r]^{d_0}
&\cE^1\ar[r]^{d_1}&\ldots
}  \label{eq:kom2}
\end{equation}
as an iteration of cones:
\begin{equation}
\Cone(\Cone(\Cone(\Cone(\ldots\to\cE^{-1})\to\cE^0)\to\cE^1)\to\ldots)
	\label{eq:cccc}
\end{equation}

In this way one may view the mapping cone as the basic operation in
the construction of complexes.

This leads one to study the structure of the category of topological
D-branes as a {\em triangulated category}. A triangle in $\DC(X)$ is
an sequence of morphisms of the form:
\begin{equation}
\xymatrix{
&C\ar[dl]^w_{[1]}&\\
A\ar[rr]^u&&B\ar[ul]^v
} \label{eq:tri1}
\end{equation}
The ``$[1]$'' denotes that the map $w$ increases the grade of any object
by one. That is, in terms of elements in the complexes, $w:c^n\to
a^{n+1}$.

There must always be exactly one edge in a triangle with a
``$[1]$''. This edge may be moved however. For example, the triangle
in (\ref{eq:tri1}) could be rewritten as
\begin{equation}
\xymatrix{
&C[-1]\ar[dl]&\\
A\ar[rr]&&B\ar[ul]_{[1]}
} \label{eq:tri2}
\end{equation}

The triangle (\ref{eq:tri1}) in $\DC(X)$ is {\em distinguished\/} if
and only if $C$ is isomorphic to $\Cone(u:A\to B)$. We refer to
chapter 5 of \cite{GM:Hom} for further details. Note that because the
``$[1]$'' can be shuffled around to any edge of the triangle, any
vertex of a distinguished triangle can be considered the mapping cone
on the opposing side of the triangle.

This democracy between the vertices of a distinguished triangle cannot
be over-empha\-sized.  Physically, it means that if two topological
D-branes $A$ and $B$ can combine to give $C$, then $A$ and $C[-1]$ can
combine to give $B$ etc. The exact meaning of this will be become more
clear when we talk about physical D-branes below.


\subsection{Physical Branes} \label{ss:phys}

We would now like to relate the topological branes above to physical
branes. 
We show in this section that many of the objects of $\DC(X)$ can
represent potential physical D-branes. In section \ref{ss:stab} we
will see that they need to satisfy further
stability requirements in order to be physical.

Let us return to the state of affairs before we introduced the derived
category and think of our D-branes as 6-branes wrapped on the \CY.
Such a brane is characterized by a choice of vector bundle $E$ over $X$:
the D$6$ charge is the rank of $E$, while the lower D-brane charges are
determined by the Chern character of $E$.  We will summarize this information
in the RR charge vector $Q_i(E)$.

This is a BPS state with a particular central change $Z(E)$. For
an even-dimensional B-brane such as this, $Z(E)$ depends purely on the
complexified K\"ahler form, $B+iJ$, and not on the complex structure
of $X$. For large radius we have \cite{MM:K,FW:D}
\begin{equation}
Z = \sum_i Q_i(E) \Pi^i 
  = \int_X e^{-B-iJ}\ch(E)\sqrt{\td(T_X)}+\hbox{quantum corrections}
	\label{eq:LRZ}
\end{equation}
We may compute $Z(E)$ exactly by passing to the mirror of $X$ as we
will see in section~\ref{ss:xi}.

The basic physical interpretation of $Z(E)$ is that $M(E)=|Z(E)|$
is the mass of the BPS brane $E$,\footnote{
This is true only if we use normalized central charges,
i.e. defined as periods of a normalized
three-form on the mirror satisfying 
$1=\int \Omega\wedge\bar\Omega$.  We will not use the
normalized masses in our subsequent considerations; clearly these are 
also valuable data (as seen in \cite{DGR:Dquin})
and it would be interesting to find some way to
use them in our considerations.
}
while the phase $\arg Z(E)$ determines
the particular $\CN=1 \subset \CN=2$ supersymmetry which is preserved by
the brane $E$.  The phase of $Z(E)$ plays many roles in the subsequent
considerations.  One of the most familiar is in defining walls of
marginal stability or ``ms-walls.''\footnote{
Of course if $\dim_\C \cM = 1$ these are lines of marginal stability,
or ms-lines.
}
Suppose we have two 6-branes $E$ and $F$ which might
combine to give a bound state $G$.  A well known argument using
$Z(G)=Z(E)+Z(F)$ and the triangle inequality shows that either
\begin{itemize}
\item $\arg Z(E) = \arg Z(F)$ implying $G$ is marginally stable for
this decay into $E$ and $F$, or
\item $\arg Z(E) \neq \arg Z(F)$ and then, assuming $G$ exists as a BPS
state, $G$ is stable against this decay.
\end{itemize}
Thus a given decay process can only take space on a specific (real)
codimension one wall in moduli space, the ms-wall, determined by the
charges $Q(E)$ and $Q(F)$.  Of course being on the wall is a necessary
but not sufficient condition for a decay.

We thus define the ``grade'' of the brane $E$ to be
\begin{equation}
\grade(E) = \frac1\pi\Im\log Z(E).  \label{eq:gradedef}
\end{equation}
While the specific $\CN=1$ supersymmetry which is preserved by $E$ is
determined by $\grade(E)$ up to shifts in $2\Z$,
we wrote $\Im\log Z$ for the phase to emphasize that (as in \cite{Doug:DC})
we define the grade to take values in $\R$, not just $[0,2)$, by
making a choice of branch for the logarithm.  

To better understand this point, consider a pair of inequivalent D-branes
whose values of $\grade$ happen to agree at a particular value of
$B+iJ$. As we vary $B+iJ$, the difference between the grades of the
branes varies continuously with $B+iJ$ but may exceed 2 even if we
remain in a fundamental region of the moduli space.  As we argue shortly,
this relative grade has physical meaning.

We therefore require a more precise definition of $\grade$ as
follows. Fix a basepoint in the moduli space $\cM$ near large radius
limit. For a given D-brane use (\ref{eq:gradedef}) to fix $\grade$ at
the basepoint in some fixed range, say, $0\leq\grade<2$. Now demand
that $\grade$ varies {\em continuously\/} along any path in the moduli
space. This makes $\grade$ a well-defined function on the {\em path
space\/} $\cP$ consisting of paths in the moduli space starting at the
fixed basepoint.

This definition fixes most of the ambiguity but not all.  First, there
is an arbitrary choice of the phase of the
holomorphic three-form (and thus all the central charges $Z(E)$) at
the basepoint.  In fact, shifting $\grade$ for all branes
simultaneously by a constant value will be physically meaningless;
only differences $\grade(E)-\grade(F)$ will enter our considerations.

Second, one wants $\grade$ to be a well-defined function on 
{\em Teichm\"uller space\/} $\cT$,
which we define as a cover of moduli space on which
central charges are single valued,
or as the moduli space of mirror quintics carrying a basis of
$H^3(X,\Z)$.
This cover should be branched only
at singular points, with branchings determined by the monodromies on
RR charges.  Now all physical quantities should be well-defined
functions on such a Teichm\"uller space, but in general $\grade(E)$ as
we have defined it may not be.  If the central charge $Z(E)$ vanishes
at a non-singular point $P$ in moduli space (this is possible if the
brane $E$ does not physically exist at $P$), paths encircling $P$ will shift
$\grade(E)$ by $\pm 2$.  We will discuss this point in
section~\ref{ss:stab}, eventually concluding that $\grade(E)$ is
well-defined only for stable objects, and that one should consider only
paths which remain in the region of stability.

The fundamental physical reason we need to lift $\grade$ to $\R$ is that 
variations of the grade (\ref{eq:gradedef}) determine the variation of the
masses of open strings stretched between branes \cite{Doug:DC}.
More precisely, the topological open strings discussed above correspond
in the physical CFT to Ramond ground states, each of which will have
a NS partner defined using spectral flow.  The $\CN=2$ superconformal algebra determines
the mass of the corresponding space-time boson in terms of its $U(1)$
charge $Q$ to be
\begin{equation}
m^2 = \half (Q-1) .
\end{equation}
We then combine this with the result that flow of grading affects the
$U(1)$ charge of a string between a brane 
$E$ and a brane $F$ as
\begin{equation}
\Delta Q = \Delta\grade(F) - \Delta\grade(E) .
\end{equation}
In particular, if $E$ and $F$ are {\em neighbors\/} in the sense that their
grades differ by one, then an element $f\in \Hom(E,F)$ corresponds
to a marginal deformation to a new state, 
the mapping cone $G=\Cone(f:E\to F)$. 

As we saw in section \ref{ss:tri}, we may use the mapping cone to
iteratively build long complexes along the lines of (\ref{eq:cccc}). 
That is one follows the following algorithm. Let us assume that the
complex begins with $\cE^0$ as the leftmost term for simplicity.
\begin{enumerate}
\item Compute $\grade(\cE^0)$ and $\grade(\cE^1)$ near large radius limit
using (\ref{eq:LRZ}). Define them to lie in some fixed range, say
$0\leq\grade<2$.
\item Let $A=\cE^0$ and $B=\cE^1$.
\item Analytically continue the $\grade$'s of $A$ and $B$ along a path in the
moduli space $\cM$ to find a point where $\grade(B)=\grade(A)+1$.
\item This defines the new object $\Cone(A\to B)$ with
$\ch=\ch(B)-\ch(A)$ and $\grade=\grade(B)$.
\item Replace $A$ by $\Cone(A\to B)$ and $B$ by the next term
in the complex.
\item Repeat from step 3 until the complex is exhausted.
\end{enumerate}
Note that we will have more to say about step 3 at the end of
section~\ref{ss:stab}. 

The property of the grading of the cone with respect to
the constituent objects we used in the rules above is justified as follows.
Referring to equation (\ref{eq:tri1}) if we have an open
string $u$ stretching from $A$ to $B$ and $\grade(B)=\grade(A)+1$, then $A$
and $B$ form a marginally bound state $C=\Cone(u:A\to B)$.
In this case $Z(C)=Z(A)+Z(B)$ where recall $A$ was viewed as an
{\em anti\/}-brane. Viewing $A$ as a brane, we have $Z(C)=Z(B)-Z(A)$, which
implies
\begin{equation}
  \grade(C) = \grade(B) = \grade(A)+1.
	\label{eq:xicone}
\end{equation}
This equation defines $\grade(C)$ and, in particular, fixes the $2\Z$
ambiguity.

In this way we can hope to build a good many objects of
$\DC(X)$. An example of a complex which {\em cannot\/} be built is
\begin{equation}
0\to\cF\to0\to\cF\to0,
\end{equation}
where $\cF$ is any sheaf. One might claim the first three terms as
giving $\Cone(\cF\to0)$ but then the relative $\grade$ of this cone and
the second $\cF$ is stuck at 2 for all values of $B+iJ$. 

This particular object is decomposable (it is a direct sum of simpler
objects) and as such would not correspond to a single physical brane;
however it is likely that many indecomposable objects in $\DC(X)$ also
cannot be constructed using the above algorithm.  $\Pi$-stability
implies that such objects never correspond to physical branes.

This definition of the grade of a complex is somewhat complicated
as it requires finding a succession of paths in moduli space, one for
each step in the procedure.  In special cases one might be able to
simplify this.  For example, one might try to build a complex
\begin{equation}
\xymatrix@1{
\ldots\ar[r]&\cE^{-1}\ar[r]^{d_{-1}}&\underline{\cE^0}\ar[r]^{d_0}
&\cE^1\ar[r]^{d_1}&\ldots
}
\end{equation}
in a single step, by finding a value for
$B+iJ$ for which $\grade(\cE^n)=n+\alpha$ for all $n$ for which 
$\cE^n$ is nonzero, and some fixed real number $\alpha$. 
However, such a value is not likely to exist if there are more
than two nontrivial terms in the complex.


\subsection{Stability and Octahedra} \label{ss:stab}

Given a point of marginal stability in the moduli space where we have
a good picture in terms the topological D-branes, we may ask what
happens if we change $B+iJ$ slightly to move away from the point.

Suppose we consider the system of $E$, $F$ and $G$ introduced above,
with $G=\Cone(f:E\to F)$, and move away from a point on the ms-wall.
Clearly one of three things can happen:
\begin{enumerate}
\item $\grade(F)-\grade(E)-1=0$ implying we remain on the ms-wall.
\item $\grade(F)-\grade(E)-1<0$ implying that the open string becomes
tachyonic. This tachyon acts like the ones studied by Sen
\cite{Sen:dbd} and binds $E$ and $F$ together to form $G$.
\item $\grade(F)-\grade(E)-1>0$.  This forces the grade of the
other two maps in the distinguished triangle to go negative, a potential
contradiction to the axioms of unitary CFT which can only be resolved if
$G$ is in fact unstable against decay
to $E$ and $F$ \cite{Doug:DC,Doug:S01}.
\end{enumerate}
Thus the grades of the D-branes determine the stability
constraints. This is the ``$\Pi$-stability'' of \cite{DFR:stab} and is
the central issue we study in this paper.

The simplest proposal one could make is that a physical D-brane is an
object in $\DC(X)$ which is $\Pi$-stable against all possible
decays. Note that $\DC(X)$ is a concept from algebraic geometry and, as
such, does not depend on $B+iJ$. The central charges do depend on
$B+iJ$ however and so therefore do the stability conditions. One might
think of the derived category data as solving some ``F-term'' equation
of motion and the further $\Pi$-stability condition as arising from
some ``D-term'' \cite{Doug:DC,Doug:S01}.

However, this proposal suffers from a problem of self-reference
which looks very problematic. In order for $C$ to
decay into $A$ and $B$, both $A$ and $B$ must actually exist! A given
triangle therefore has little information content:
\begin{itemize}
\item If $\grade(B)-\grade(A)<1$ we cannot assert that $C$ is stable as it
may be the vertex of another triangle which allows a decay into other
products. 
\item If $\grade(B)-\grade(A)>1$ we cannot assert that $C$ is unstable as we
have not established the stability of $A$ or $B$.
\end{itemize}

Conceptually, the simplest way to deal with this problem is to assume
that there is some other way of determining the set of stable objects
at a given ``basepoint'' $P_0\in\cM$ and then try to follow the set of
stable objects as we move away from this basepoint. We define
$\Pi$-stability as a criterion on the entire set of stable objects
\cite{Doug:DC}.  This set or list, call it $S(P)$, should be
considered as a function on path space.  We furthermore assign a value
of $\grade$ (fixing the mod 2 ambiguity) to each object in $S(P)$.

The rules above can now be used to determine the entire list $S(P')$
at all nearby points $P'$ given $S(P)$.  To be more precise, we enumerate
the following constraints, to be applied at every point $P$:
\begin{enumerate}
\item 
For every distinguished triangle of the form 
\begin{equation}
\xymatrix{
&C\ar[dl]_{[1]}&\\
A\ar[rr]&&B\ar[ul]
} \label{eq:tri10}
\end{equation}
with $A,B\in S(P)$, $\grade(B)-\grade(A)=1$, we have $C\in S(P)$ with
$\grade(C)=\grade(B)$.
\item
On the side $\grade(B)-\grade(A)<1$ of the ms-wall defined
in 1, we have $C\in S(P)$, with all grades $\grade(A)$, $\grade(B)$ and
$\grade(C)$ determined by continuity.
\item
Conversely, for every triangle of the form (\ref{eq:tri10}) with $A,B \in S(P)$
and $\grade(B)-\grade(A)>1$, we have $C\not\in S(P)$.
\end{enumerate}

To get the definition started, one starts at a base point $P_0$ in
moduli space where $S(P)$ is known {\it a priori}. For example, in
the large volume limit, one might take $S(P)$ to be the $\mu$-stable
coherent sheaves.\footnote{ 
Strictly speaking, this will be true only
in the limit in which the volume (in string units) is much larger than
any other quantity in the problem at hand (in particular, the charges
$Q$ of the objects under discussion).  It is also possible that one
must make some restriction on the types of singularities allowed in
these sheaves.  
} 
One might start from other points as well; an
example we will discuss in more detail below is the orbifold point.

It is a very interesting question as to whether $\Pi$-stability can
intrinsically determine the set of stable objects without need for
other criteria. We will not attempt to address this problem in this
paper.

In the examples, $S(P)$ is an infinite set, so one needs further
information to use this definition in practice.  For now, let us start
by formulating the criterion in a way which is clearly practical,
i.e. could be implemented on a computer.

Thus, let $Q$ be a subcategory of $\DC(X)$ consisting of a finite
number of objects together with a finite number of morphisms.  There
are therefore a finite number of distinguished triangles composed
purely of objects and morphisms in $Q$.  This allows explicitly
checking the rules above, and objects in $S(P)$ constructed in this
way will be referred to as ``$Q$-stable''.  Hopefully as $Q$ is taken
to be a larger and larger subset of $\DC(X)$ we get a closer
approximation to the truth of whether the corresponding D-brane
really exists as a physical object. There are probably many
mathematical subtleties which need to be overcome to make this hope
more manifest.

The first criticism one can make of this proposal is that stability,
which should be a physical notion, depends on position in the path
space $\cP$ rather than the moduli space itself. As we explore in
section \ref{s:mon}, the physical identity of an object in $\DC(X)$
undergoes monodromy on going around singularities in $\cM$ and so one
cannot expect the notion of stability of an object in $\DC(X)$ to
depend only on the position in $\cM$. However, one should expect stability to
be a function of position in the Teichm\"uller space $\cT$.
That is, a loop in $\cM$ not enclosing anything like a conifold
point, Gepner point, etc., should be trivial and so the notion of
stability should not change upon monodromy around this loop.

As we commented earlier, the value of $\grade$ for objects with
general charge will be ambiguous on $\cT$.  However, our rules for
stability only use the $\grade$'s of stable objects; indeed there is
no clear physical definition of $\grade$ for an unstable object.

Grades of stable objects will be unambiguous on $\cT$ 
if no path in $\cR(A) \subset \cT$, the region in which $A$ is stable,
encloses a zero $Z(A;P)=0$.
This will be true if we grant
two conditions.  First, stable objects cannot become massless at
non-singular points in moduli space,
\begin{equation} \label{eq:massconsist}
Z(A;P) \ne 0 \qquad \forall A \in S(P) .
\end{equation}
Of course this is also a physical consistency condition --- such a
massless state would have shown up as a singularity of the conformal field
theory and in the conjugate period.  Second, we will assume that
$\cR(A)$ is simply connected.\footnote{
This is true in all known examples, but we should say that it is not
completely obvious and it would be quite interesting if a
counterexample were found.
}

In the finite approximation of $Q$-stability, a massless object in $Q$ which
is $Q$-stable at a smooth point in $\cM$ should become $Q$-unstable if
we increase the size of $Q$ to more closely approximate real physics.
We therefore define a collection $Q$ to be ``$s$-complete'' if no
$Q$-stable objects become massless at a (smooth) point in $\cM$.

\iffigs
\begin{figure}
  \centerline{\epsfxsize=8cm\epsfbox{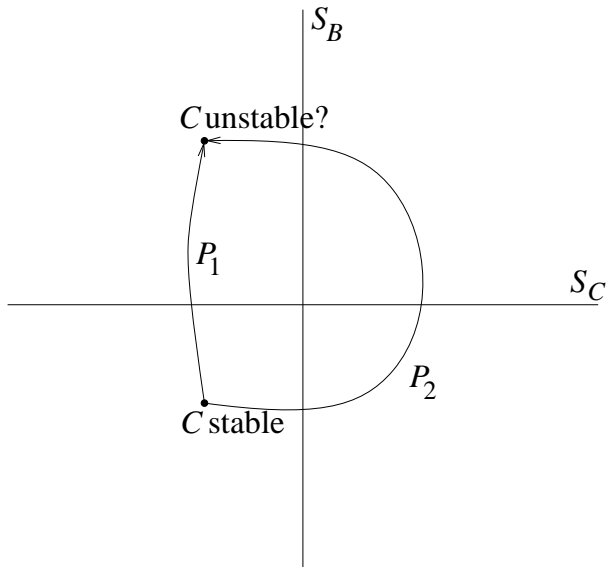}}
  \caption{Competing lines of stability.}
  \label{f:2stab}
\end{figure}
\fi

We now argue that $s$-completeness is sufficient to guarantee that the
notion of $Q$-stability is well-defined on $\cT$. Let us consider an
object $C$ which we declare to be stable at a given point in the path
space $\cP$. One possible decay path of $C$ is $C\to A+B$. In
figure~\ref{f:2stab} we denote by $S_C$ a codimension one wall in $\cT$ with
$\grade(B)-\grade(A)=1$. This is an ms-wall
if $A$ and $B$ are stable. 

Suppose there is also the
possibility of $B$ decaying into $E+F$. We denote the corresponding
ms-wall by $S_B$. Assume $B$ is stable to the left of this line.
Let us ignore all other possible triangles (i.e.,
decays). Following path $P_1$ in the figure we see that $C$ decays as
we cross $S_C$. Thus the object $C$ is unstable at the end of the
path.

Consider now the path $P_2$ in figure~\ref{f:2stab}. At the time it
crosses $S_C$, the object $B$ is no longer stable and so we cannot
claim $C$ necessarily decays. This could lead to the unpleasant
conclusion that $C$ is still stable if we follow the path $P_2$. This
is exactly the situation we need to avoid in order for stability to
depend only on the position in $\cT$, since $\cT$ essentially represents
{\em homotopy classes\/} of paths.

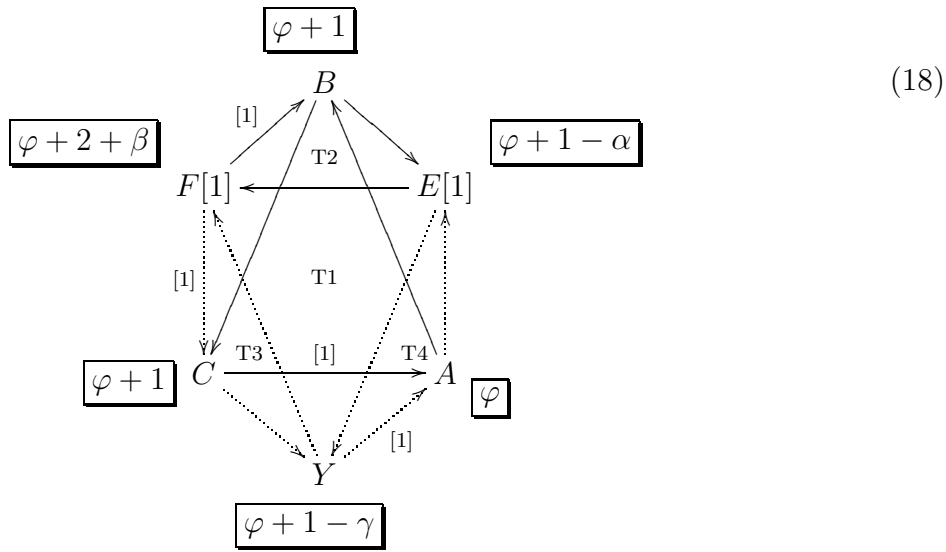
\begin{figure}[t]
\begin{equation}
\begin{xy}
\xymatrix{
&B\ar[lddd]\ar[rd]&\\
F[1]\ar@{.>}[dd]_{[1]}\ar[ur]^{[1]}&&E[1]
        \ar[ll]\ar@{.>}[dddl]\\
&&\\
C\ar[rr]^{[1]}\ar@{.>}[dr]&&A\ar[uuul]\ar@{.>}[uu]\\
&{Y}\ar@{.>}[ur]_{[1]}\ar@{.>}[uuul]&
}
\POS<38mm,-42mm>*+<6pt>{\grade}\drop\frm{-,}
\POS<48mm,-8mm>*+<6pt>{\grade+1-\alpha}\drop\frm{-,}
\POS<14mm,-59mm>*+<6pt>{\grade+1-\gamma}\drop\frm{-,}
\POS<-10mm,-40mm>*+<6pt>{\grade+1}\drop\frm{-,}
\POS<-16mm,-8mm>*+<6pt>{\grade+2+\beta}\drop\frm{-,}
\POS<14mm,7mm>*+<6pt>{\grade+1}\drop\frm{-,}
\POS<16mm,-26mm>*{\txt{\scriptsize T1}}
\POS<16mm,-10mm>*{\txt{\scriptsize T2}}
\POS<6mm,-36mm>*{\txt{\scriptsize T3}}
\POS<28mm,-36mm>*{\txt{\scriptsize T4}}
\end{xy}
\end{equation}
\caption{The octahedron for $C=A+B$ and $B=E+F$.} 
\label{f:oct1}
\end{figure}

Fortunately the ``octahedral axiom'' rescues us. For details of the
octahedral axiom we refer to \cite{GM:Hom}.  This axiom,
which is obeyed by $\DC(X)$, states a general
structure associated to two distinguished triangles with a common
vertex.  To resolve the problem, it must be that in this situation,
there always exists a third triangle predicting that if $B$ decays,
$C$ can still decay into branes including the decay products of $B$.
This is just what we will find.

Suppose $A$ and $B$ marginally bind to form $C$ and suppose $B$ itself
is a marginally bound state of $E$ and $F$.
The two distinguished triangles are shown by solid arrows in figure
\ref{f:oct1}. In order to get the ``[1]''s at a convenient edge
$E$ and $F$ are replaced by $E[1]$ and $F[1]$. The octahedral axiom
states that given such a pair of triangles
forming the ``upper half'' of an octahedron, there exists an object
$Y$, unique up to isomorphism, such that two more distinguished
triangles exist as shown by dotted arrows in figure \ref{f:oct1}. 
The octahedron thus consists of four distinguished triangles labeled
at their centre by T1\ldots T4 in figure \ref{f:oct1}. 
The four remaining faces of the octahedron commute. 
That is, two dotted arrows compose to give an undotted arrow etc.

Let $\grade$ denote $\grade(A)$. One may then use (\ref{eq:xicone}) to
compute the grades of the remaining 5 objects in the octahedron.
These grades are shown (with $\alpha=\beta=\gamma=0$) in the boxes in figure
\ref{f:oct1}.

The triangle T3 in figure \ref{f:oct1} thus shows that $C$ is
marginally stable with respect to a decay into $F$ and $Y$ (in
addition to T1 giving a decay into $A$ and $B$). Triangle T4 shows
that $Y$ itself is marginally stable with respect to a decay into $A$
and $E$.

Now let us change $B+iJ$ a little to change some stability constraints.
Suppose we preserve the marginal stability of T1 giving $C=A+B$
but we destabilize $B$ so that it decays into $E+F$.
This implies that we have arranged $\grade(F)-\grade(E)>1$. Because 
$Z(B)=Z(E[1])-Z(F[1])$, $\arg Z(B)$ must lie between $\arg(E[1])$ and 
$\arg(-F[1])$. In other words if we do not change $\grade(A)$, and thus
$\grade(B)$, the grade of $F$ must increase a little and the grade of 
$E$ must decrease a little to make $\grade(F)-\grade(E)>1$. Similarly the
grade of $Y$ will decrease a little.

We show the resulting new grades in figure \ref{f:oct1} where
$\alpha$, $\beta$ and $\gamma$ are all positive numbers. This has an
interesting effect on the triangles T3 and T4. First note that T4
gives $\grade(E[1])-\grade(A)<1$ which implies that $Y$ is now {\em
stable}. Second T3 gives $\grade(F[1])-\grade(Y)>1$ which implies that
$C$ is {\em unstable}.

We have therefore shown that if $C$ is marginally stable against decay
into $A$ and $B$, but that $B$ itself is unstable with respect to a
decay into $E$ and $F$, then $C$ will always be unstable with respect
to decay into $F$ and some bound state of $A$ and $E$.

Referring back to figure~\ref{f:2stab} we have shown that at the time
the path $P_2$ crosses the line $S_C$, the object $C$ must have
already decayed into something else. In other words there must be
another wall of marginal stability encountered before $S_C$. We
therefore have the consistent result that $C$ is unstable at the end
of the path in figure~\ref{f:2stab} independent of the homotopy
class of the path.

We give an example of this effect in section \ref{ss:X2} which we hope
will clarify the argument above.


\subsection{Stable Sheaves and Vector Bundles} \label{ss:bun}

It is well worth comparing the notion of stability for D-branes with
similar notions for vector bundles and sheaves. 
The familiar notion of a D-brane, before venturing into the
complexities of the derived category, is that of a vector bundle $E$
supported on $X$ on over some subspace of $X$. The connection on this
bundle must satisfy the Hermitian-Yang-Mills equation. 

A bundle $E$ has a ``slope'' $\mu$ given by
\begin{equation}
\mu(E) = \frac{1}{\rank(E)} \int c_1(E)\wedge (*J)
\end{equation}
where $J$ is the K\"ahler form.
A bundle $E$ is said to be {\em stable\/} if it is associated to a
sheaf $\cE$ such that every coherent
subsheaf $\cF$ of lower rank satisfies $\mu(\cF)<\mu(\cE)$. It is
{\em semistable\/} if the weaker condition $\mu(\cF)\le\mu(\cE)$ holds.

We then have the following theorem due to Donaldson, Uhlenbeck and Yau
\cite{Don:YM,UY:YM}
\begin{theorem}
An bundle is stable if and only if it
admits an irreducible Hermitian-Yang-Mills connection. This connection
is unique.
\end{theorem}

As shown in \cite{DFR:stab}
this stability condition is implied by $\Pi$-stability
at large radius limit as we now review. Suppose $\cF$ is a subsheaf of
$\cE$. We then 
have a short exact sequence
\begin{equation}
 0 \to \cF \to \cE \to \cG \to 0,  \label{eq:FEG}
\end{equation}
for some $\cG$ which yields a distinguished triangle in the derived
category in the usual way. Near the large radius limit we may
use~(\ref{eq:LRZ}) to show
\begin{equation}
 \grade(\cE) \approx -\frac1\pi\tan^{-1}\frac{2V}{\mu(\cE)},
\end{equation}
where $V$ is the volume of $X$.  The stability condition
$\mu(\cF)<\mu(\cE)$ is then equivalent to
$\grade(\cF)<\grade(\cE)$. Since $\ch(\cE)=\ch(\cF)+\ch(\cG)$, we have
$Z(\cE)=Z(\cF)+Z(\cG)$ and so $\grade(\cE)$ lies between $\grade(\cF)$
and $\grade(\cG)$. It follows that $\mu(\cF)<\mu(\cE)$ implies that
$\grade(\cF)-\grade(\cG[-1])<1$ and so $\cE=\Cone(\cG[-1]\to\cF)$ is
$\Pi$-stable with respect to the triangle associated to
(\ref{eq:FEG}).

Comparing this with the earlier discussion of $\Pi$-stability, the
main point of difference is that the coherent sheaves form an
{\em abelian\/} category: one has a well-defined notion of
kernels and cokernels. The statement that ``$\cF$ is a subsheaf of
$\cE$'' is another way of saying that the morphism $\cF \to \cE$ has
no kernel and so the statement depends on the abelian structure of the
category. 

The derived category of coherent sheaves is not an abelian category
and hence has no absolute notion of ``sub-D-brane''. In checking the
stability of a sheaf one has to check for stability only against all
subsheaves. Since a subsheaf has a lower rank, this is always a finite
process. For D-branes however there is no such hierarchy of objects
such as rank and therefore no analogue of a finite chain of possible
decay products. Given two
D-branes $A$ and $B$, it is known in examples that $A$ can decay into
$B$ for some values of the K\"ahler form, while $B$ can decay into
$A$ for other values, as we will see in subsection 4.1.
It is this lack of hierarchy between D-branes and their decay products
which forces one to use a category which is not abelian, rather than
simpler notions such as sheaves.

\subsection{Abelian subcategories and the orbifold point}

The arguments we just made do not yet rule out the possibility that
in {\em specific regions\/} of Teichm\"uller space, one can reduce the general
discussion of $\Pi$-stability to stability in an abelian category,
as proposed in \cite{DFR:stab}.
Thus, let us explore the claim that in some region $\cR$, the
$\Pi$-stable objects are precisely those objects $E$
which satisfy $\grade(E')<\grade(E)$ for all subobjects $E'$ of $E$
defined with respect to an abelian category $\cA_\cR$.

It is convenient to impose the further requirement that $\cA_\cR$ is a
graded abelian category with integer gradings (as usual).  

At present,
the only known case of a point at which all relative gradings are
integer, $\grade(E)-\grade(F) \in \Z$, is the orbifold point (for a
$\C^3/\Gamma$ orbifold singularity).  Near such an orbifold point, we can
use conventional $\CN=1$ supersymmetric gauge theory to describe
all bound states of fractional branes, as has been observed in
many works.  We now give physical arguments why this is so, to better
understand in what situations this simplification might hold.

Consider a combination of fractional branes; 
it will fall into one of three classes.  
If the central charges of all the branes
roughly align, then all tachyonic modes have small
masses compared to the string scale, and the conditions for applicability
of low energy world-volume field theory are clearly met.
On the other hand, if some branes have roughly anti-aligned central
charge, we can further distinguish the cases in which a fractional 
brane and its own antibrane are both present, and the case in which
this is not so.  In the first case, since the mass of the brane-antibrane
tachyon is string scale and much larger than any other scale, the lowest
energy configuration clearly will condense this tachyon, leading to
brane-antibrane annihilation and a configuration with a simpler description
(in terms of fewer fractional branes).  In the second case, examination of
the orbifold spectrum shows that there are no brane-antibrane tachyons
present, so no possibility for a bound state involving all of the objects.
Again, the simple bound states can be described using only branes or only
antibranes.

This last point in the argument relies on a condition on the spectrum
which holds near the orbifold point but not necessarily for a general
collection of branes.  Thus the condition that the constituent brane
central charges roughly align is necessary but not sufficient for the
applicability of supersymmetric gauge theory.  The additional
condition that the only string-scale tachyons are those between a
brane and its own antibrane is sufficient; we have not established
that it is necessary.

Having justified the use of supersymmetric gauge theory in a region $\cR_{orb}$
around the orbifold point, we note that indeed $\Pi$-stability reduces
to a stability condition in an abelian category in this region; in fact
we have justified the use of $\theta$-stability near the orbifold point
as in \cite{DFR:orbifold},
using the category of quiver representations.

We now ask what is the boundary of this region $\cR_{orb}$.
The primary constraint is that morphisms $\Hom(E,F)$ of negative
degree can only be produced by flow from morphisms of degree zero.
This is clear because we require every decay predicted by
$\Pi$-stability to correspond to a decay into a subobject.

A sufficient (but perhaps not necessary) condition for this
is to require that, compared to the orbifold point,
\begin{equation}
|\Delta\grade(E)-\Delta\grade(F)| < 1
\end{equation}
for all differences of gradings of objects in $\cA_\cR$.

This condition determines the region $\cR_{orb}$ around the orbifold
point in which the category of quiver representations can be used to
define stability.  For $\C^3/\Z_3$, its boundary (within the fundamental
region) is the ms-line on which the D$2$-brane becomes stable.
It is easily seen that this stability condition breaks down on the
boundary of this region.

Within $\cR_{orb}$, the appropriate notion of stability is
a generalized notion of $\theta$-stability, which uses the same definition
of subobject, but in which the stability
condition $\theta \cdot n > \theta \cdot n'$ for every subobject
$E'\to E$ is replaced by the condition $\grade(E') < \grade(E)$.
The hierarchy between branes and decay products is governed by the
total number of fractional branes, which is always decreasing in subobjects;
this number is determined by the charge of an object.

For more general (compact) \CY's, there probably does not exist any point
in moduli space at which all brane central charges roughly align or
antialign.  Such points will exist for subcategories of branes
carrying a subset of the possible charges, and it would probably be quite
useful to analyze the spectrum within these subcategories along the
lines which we just described.  Let us however continue to discuss proposals
which could describe all the branes.

A logical extension of the considerations in this subsection is to try
to find a set of regions $\cR_i$ whose union covers the Teichm\"uller
space, and corresponding graded categories $\cA_i$ each appropriate in
a region $\cR_i$, meaning that within $\cR_i$ all decays predicted by
$\Pi$-stability in fact reduce to decays into subobjects as defined
within $\cA_i$.  If such a picture can be made to work, it would be
reasonable to refer to the regions $\cR_i$ as ``phases,'' as the choice
of $\cA_i$ already tells us quite a lot of information about the BPS
spectrum.

The simplest proposal for constructing such an abelian category
$\cA_i$ goes as follows.  First, one needs to postulate integer
gradings for all of the objects.  The simplest prescription one can
make is to take the gradings $\grade(E)$ at a specific point $P$ in
moduli space and round them off to the nearest integer; call this
$n(E)$.  As long as we make a consistent choice of $n(E)$ for each
$E$, any choice of rounding will work.  This choice then leads to a
definition of integer grading for each morphism; let $n(f)$ be the
grading of the morphism $f$.

As in the orbifold example, we want to arrange things so that all
decays predicted by $\Pi$-stability in fact proceed through subobjects
as defined by exact sequences involving morphisms of grading $n=0$.
This condition will again define the region $\cR$ in which a given choice
of integer gradings can be used.  A potential difficulty with this idea
appears at this point, as there is no guarantee that the region $\cR$ so
defined will have finite extent.

Assuming it does, one then wants to show that this data indeed defines
an abelian category; in particular that morphisms have kernels and
cokernels.  As mentioned in \cite{Doug:DC}, the appropriate framework
for this discussion is the formalism of $t$-structures; we hope to return
to this in future work.


\section{Examples on the Quintic}   \label{s:eg}

The analysis of section~\ref{s:pistab} is clarified in this section by
analyzing various examples of $\Pi$-stability in the context of the
quintic hypersurface in $\P^4$. 


\subsection{Basics of the moduli space and periods} \label{ss:quin}

In this section we review the techniques required from Candelas et al
\cite{CDGP:}. Let $X$ be the quintic hypersurface in $\P^4$. The
mirror $Y$ is then a $(\Z_5)^3$ orbifold of a quintic with general
defining equation
\begin{equation}
  x_0^5+x_1^5+x_2^5+x_3^5+x_4^5-5\psi x_0x_1x_2x_3x_4.
\end{equation}
The single complex variable $\psi$ spans the one-dimensional moduli
space of complex structures of $Y$. This encodes the mirror of the
one-dimensional moduli space of complexified K\"ahler forms on $X$.
It is useful to introduce the variable $z=(5\psi)^{-5}$.

The Gepner point of the moduli space of $X$ is mirror to $\psi=0$. The
large radius limit of $X$ is mirror to $z=0$. At $\psi=1$, one has
the ``conifold'' point where $Y$ becomes singular (acquiring a
conifold point). This is mirror to the fact that the conformal field
theory on $X$ becomes singular for the corresponding value of $B+iJ$.

Since $\dim H_3(Y)=4$, we have 4 independent periods $\int_\Gamma
\Omega$ for the holomorphic 3-form $\Omega$. These periods satisfy the
Picard-Fuchs equation (see \cite{Mor:PF} for a nice account of this).
Following \cite{CDGP:} we use the following periods given as a power
series around the Gepner point:
\begin{equation}
 \varpi_j = -\ff15\sum_{m=1}^\infty\frac{\alpha^{(2+j)m}\Gamma(\frac m5)}
  {\Gamma(m)\Gamma(1-\ff m5)^4}z^{-\frac m5}.  \label{eq:varpi}
\end{equation}
We may use
$\varpi_0,\ldots,\varpi_3$ as a basis.\footnote{
In this section and section 4, we follow the convention that the
relation (\ref{eq:gradedef}) between $\varphi(E)$ and $Z(E)$ has
an additional minus sign, and $Z(E)$ is the complex conjugate of that 
defined in (\ref{eq:LRZ}).
}

One may then analytically continue these periods using Barnes'
integral methods to match this basis with series around the
large radius limit. We use the following basis near $z=0$:
\begin{equation}
\begin{split}
\Phi_0 &= \frac1{2\pi i}\int\frac{\Gamma(5s+1)\Gamma(-s)}
  {\Gamma(s+1)^4}(e^{\pi i}z)^s\,ds\\
 &= \sum_{n=0}^\infty\frac{(5n)!}{n!^5}z^n\\
 &= -\ff15\sum_{m=1}^\infty\frac{\alpha^{2m}\Gamma(\frac m5)}
  {\Gamma(m)\Gamma(1-\ff m5)^4}z^{-\frac m5}\\
 &= \varpi_0.
\end{split}
\end{equation}

\begin{equation}
\begin{split}
\Phi_1 &= -\frac1{2\pi i}\cdot\frac1{2\pi i}
  \int\frac{\Gamma(5s+1)\Gamma(-s)^2}
  {\Gamma(s+1)^3}z^s\,ds\\
  &= \frac1{2\pi i}\log z + O(z) = t+O(e^{2\pi it})\\
  &= -\ff15(\varpi_0-3\varpi_1-2\varpi_2-\varpi_3).
\end{split} \label{eq:Phi1}
\end{equation}

\begin{equation}
\begin{split}
\Phi_2 &= -\frac1{2\pi^2}\cdot\frac1{2\pi i}
  \int\frac{\Gamma(5s+1)\Gamma(-s)^3}
  {\Gamma(s+1)^2}(e^{\pi i}z)^s\,ds\\
 &= -\frac1{4\pi^2}(\log z)^2+\frac1{2\pi i}\log z-\ff56+O(z)\\
 &= t^2+t-\ff56+O(e^{2\pi it})\\
 &= \ff25(-2\varpi_0+\varpi_2+\varpi_3).
\end{split}
\end{equation}

\begin{equation}
\begin{split}
\Phi_3 &= \frac1{(2\pi i)^3}\cdot(-6)\cdot\frac1{2\pi i}
  \int\frac{\Gamma(5s+1)\Gamma(-s)^4}
  {\Gamma(s+1)}z^s\,ds\\
 &= \frac i{8\pi^3}(\log z)^3 + \frac{7i}{4\pi}\log z-
  \frac{30i\zeta(3)}{\pi^3} + O(z)\\
 &= t^3-\ff72t - \frac{30i\zeta(3)}{\pi^3} + O(e^{2\pi it})\\
 &= -\ff65(2\varpi_1+2\varpi_2+\varpi_3).
\end{split} \label{eq:Phi3}
\end{equation}

These analytic continuations are valid for
\begin{equation}
  -2\pi < \arg z < 0.  \label{eq:argz}
\end{equation}

In each case above the integral is taken along the path corresponding
to a straight line from $s=\epsilon-i\infty$ to $s=\epsilon+i\infty$,
where $0<\epsilon<1$. The variable $t$ above corresponds to $B+iJ=te_1$
for $X$ where $e_1$ generates $H^2(X,\Z)$. The ``mirror map'' determines
$t$ uniquely as
\begin{equation} \label{eq:deft}
  t = \frac{\Phi_1}{\Phi_0}.
\end{equation}
We refer to \cite{CK:mbook}, for example, for an account of the
history and status of the mirror map. 

The above results show that the true stringy complexified K\"ahler
moduli space $\cM$ of $X$ is the $\P^1$ with affine coordinate $z$; the
periods are singular or have branch points at the three points
mentioned above. (The Gepner 
point is only an orbifold singularity, and physics is nonsingular
there.)  

As discussed in section~\ref{ss:stab} the stability questions are best
thought of in terms of the Teichm\"uller space rather than the moduli
space. In terms of the ``presentation of data'' we wish to give in
this example, we basically have three choices of how to parametrize the
K\"ahler moduli space:
\begin{enumerate}
\item We can use $z$ (or $\psi$) as above as a parameter on the moduli
space. Since $z$ is not a parameter on the Teichm\"uller space, the
periods and questions of stability will not be a single valued
function of $z$.
\item We can use the theory of automorphic functions to construct the
Teichm\"uller space as the upper half-plane as was done in section 2.2
of \cite{CDGP:}. This has the advantage of presenting the periods as
single valued functions. The disadvantage of this approach is that
this Teichm\"uller space appears to be remarkably unphysical. For
example, the real line of the upper half-plane will contain a
dense set of copies of both the conifold point and the large radius
limit point. The former are a finite distance from a generic point in
the moduli space while the latter are an infinite distance! 
This is tied in with problems of notions of T-duality
for such string compactifications \cite{AP:T}.
\item We can use $t=B+iJ$ as the parameter. This is the most physical
parameter for the moduli space. The disadvantage of this approach is
that fundamental regions of the moduli space do not ``tile'' the
$t$-plane leading to multivaluedness as in the $z$-plane. This
problem should be clear from the figures in section~\ref{s:eg}.
\end{enumerate}

The choice we make in this paper is to go with the third
option and use $t$ to plot lines of marginal stability. This is
viable because the lines of marginal stability we consider
usually remain in the ``\CY'' phase, i.e., remain on the large
radius side of the conifold point. In this phase, $t$ is certainly the
most natural parameter. Some of the ms-lines analyzed later in this section
venture outside this 
phase but they do not do unpleasant things such as crossing
themselves when plotted in the $t$-plane. If such pathologies occur,
as they surely do in more 
complicated examples, it might be easier to
visualize the ms-lines in the less physical choice of using
the upper half-plane as the Teichm\"uller space.


\subsection{The exact computation of $\grade$} \label{ss:xi}

For a \CY\ threefold, an A-type D-brane corresponds to a 3-cycle. The
central charge, $Z$, of this D-brane is then given, up to some
normalization, exactly by the period
associated to this cycle. The normalization will not concern us
since we are only interested in differences in gradings and hence
ratios of periods.
We may use this fact, together with the
knowledge of section \ref{ss:quin}, to determine the exact central
charges of B-type branes on the quintic threefold.

Given a particular element of $\DC(X)$ we may compute the Chern
character using
\begin{equation}
\ch(\cE^\bullet) = \sum_n(-1)^n\ch(\cE^n).
\end{equation}
Equation~(\ref{eq:LRZ}) then determines the large radius behaviour of
$Z(\cE^\bullet)$ as a series in $t$. This is then enough to fix the
exact linear 
combination of $\varpi_j$'s from section \ref{ss:quin} which
represents the period on $Y$ and hence the exact form of
$Z(\cE^\bullet)$.

In order to perform this analysis we need to know precisely what
constitutes the ``quantum corrections'' of (\ref{eq:LRZ}).
These corrections come from two sources:
\begin{enumerate}
\item Nonperturbative corrections: These are of the form $a_n\exp(2\pi
int)$ for $n>0$.
\item Perturbative corrections: These are of the form $b_m t^m$. Since
these corrections begin at ``four loops'' \cite{GVZ:4loop}, the
highest power of $t$ appearing will be three less than the leading
order (``one loop'') term.
\end{enumerate}
Note therefore that the only effect of the perturbative corrections is to
contribute a constant term in (\ref{eq:Phi3}). We claim that this
constant term acts to cancel the ${30i\zeta(3)}/{\pi^3}$ term in
(\ref{eq:Phi3}). One way to justify this is to do a direct computation
of the 4-loop term which we have not done. 

Instead let us use the fact the the singularity in the conformal field
theory associate to $X$ at the ``conifold point'' is associated to a
D-brane becoming massless. The monodromy of the periods around the
conifold point given by table 3.1 of \cite{CDGP:} immediately implies
that the period associated to this D-brane must be
$-\varpi_0+\varpi_1$. The analysis of the previous section shows that
at large radius this period behaves as
\begin{equation}
\ff56t^3+\ff{25}{12}t-\frac{25i\zeta(3)}{\pi^3}.
\end{equation}
Using (\ref{eq:LRZ}) and $\td(T_X)=1+\ff56 e_1^2$ for the quintic we have
\begin{equation}
  \ch(\cE^\bullet) = 1 + \left(c-\frac{25i\zeta(3)}{\pi^3}\right)e_1^3,
\end{equation}
where $c$ represents the 4-loop correction. Since $c$ is imaginary
(see \cite{CDGP:} for example) the reality of $\ch$ confirms our
assertion.

Furthermore we see immediately that the D-brane that vanishes at the
conifold point satisfies $\ch(\cE^\bullet) = 1$. It is therefore
natural to associate it with the structure sheaf $\O$ as asserted in 
\cite{BDLR:Dq}.

Now that we have fixed this 4-loop term we have an exact method for
associating a linear combination of $\varpi_j$'s (over the rational
numbers) with the central charge of any element of $\DC(X)$. We
therefore have an exact method of computing the grade $\grade$ of any
B-type D-brane on the quintic. This method should generalize to any
example where the Picard-Fuchs system is understood in the language
of generalized hypergeometric functions.

In the examples below we computed $\varphi$ numerically by integrating
the Picard--Fuchs equation with a Runge--Kutta algorithm. The initial
conditions were specified by using the power series expansions of
$\varpi_j$ (\ref{eq:varpi}).


\subsection{$N$ D4-branes}  \label{ss:D4}

As a first example consider the following map
\begin{equation}
\xymatrix@1{
&\O(-N)\ar[r]^-{f(x_i)}&\O,
} \label{eq:D4-N}
\end{equation}
where $f(x_i)$ is a polynomial of degree $N$ in the homogeneous
coordinates of $\P^4$. By $\O(-N)$ we mean the restriction of the line
bundle on $\P^4$ with $c_1=-N$ to the quintic $X$.
If $N=1$ then the short exact sequence
\begin{equation}
\xymatrix@1{
0\ar[r]&\O(-1)\ar[r]^-{f(x_i)}&\O\ar[r]&\O_D\ar[r]&0,
}
\end{equation}
implies that the cone of the map (\ref{eq:D4-N}) is simply
$\O_D$. Here $D$ is the divisor 4-cycle in $X$ given by the zeroes of
$f(x_i)$ and $\O_D$ is its structure sheaf extended by zero over
$X$. That is $\O_D$ is the sheaf representing the 4-brane wrapped on
$D$.

Similarly for general $N$ we have a collection of $N$ 4-branes on
$X$. If $f(x_i)$ takes on particular forms we may make these 4-branes coalesce
or not as we wish. 

The methods of section~\ref{ss:xi} can be used to determine
\begin{equation}
  Z(\O(-N)) = -\ff16(5N^3+3N^2+16N+6)\varpi_0
     +\ff12(3N^2+3N+2)\varpi_1
     +N^2\varpi_2 +\ff12N(N-1)\varpi_3.
\end{equation}

\iffigs
\begin{figure}[t]
  \centerline{\epsfxsize=15cm\epsfbox{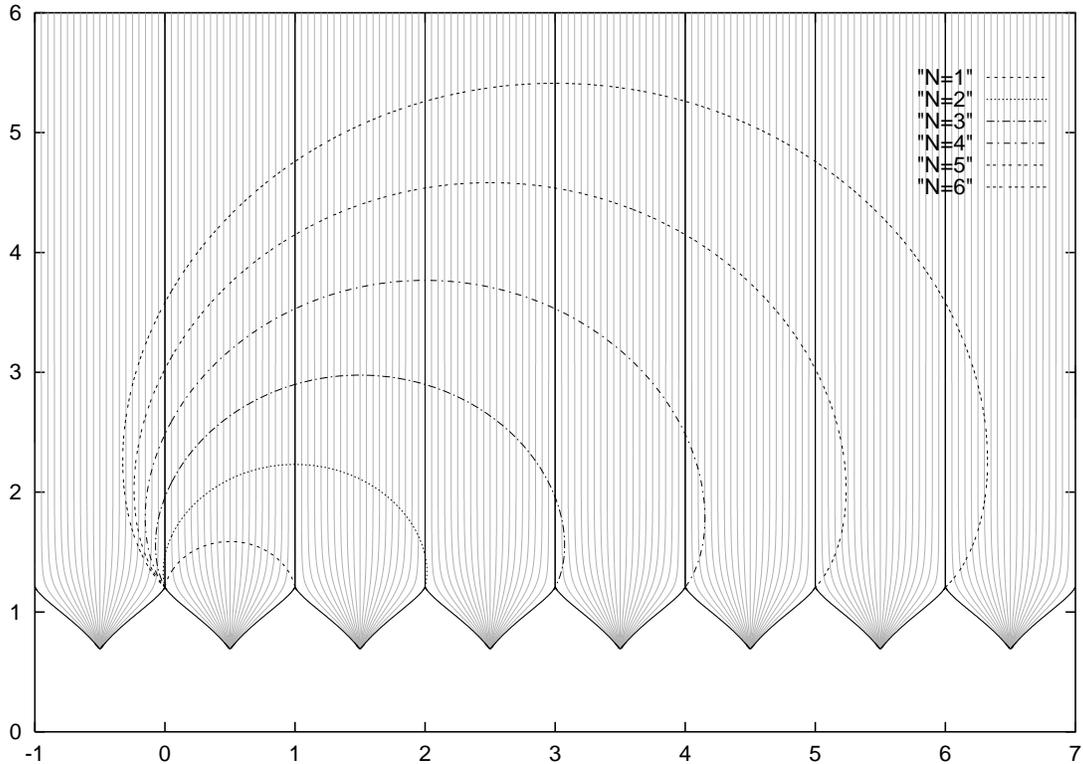}}
  \caption{Stability of $N$ D4-branes in the $t$-plane.}
  \label{f:D4}
\end{figure}
\fi

A collection of $N$ 4-branes is stable against decay to
the 6-brane $\O$ and the anti-6-brane with 4 brane charge $-N$ if
\begin{equation}
  \grade(\O)-\grade(\O(-N)) < 1.
\end{equation}
The resulting lines of marginal stability for various values of $N$
are shown in figure~\ref{f:D4}. The D4 branes are stable {\em above\/}
the line (i.e., large radius) as one would expect.
Note that this picture is of a series
of fundamental regions in the $t$-plane each spanning a shift in the
$B$-field (i.e., real part of $t$) by one. The starting region, as
required by (\ref{eq:argz}), is the one where $-1<B<0$. The regions to
the right are obtained by analytic continuation.

Note that the lines of marginal stability have one end at the conifold point
where $Z(\O)$ vanishes and the other end is at the copy of the conifold
point where $Z(\O(-N))$ vanishes. This is a general feature of the
lines of marginal stability. They can only end
at a point where $\arg(Z)$ becomes ill-defined, i.e., where one of
the periods vanishes.

Note that for large $N$ one may use (\ref{eq:LRZ}) directly to show
that asymptotically the line of stability becomes an arc of a circle
of radius $N/\sqrt{3}$ with centre $t=N/2 + iN/2\sqrt{3}$.


\subsection{Some exotic D-branes}  \label{s:exot}

As discussed in \cite{Doug:S01} one may use the D4-brane construction
in section~\ref{ss:D4} as a basis of a construction of more exotic
objects. The map $f(x_i)$ in (\ref{eq:D4-N}) lives in the space
$\Hom(\O(-N),\O)=\Hom(\O,\O(N))=\Ext^0(\O,\O(N))=H^0(\O(N))$ (see, for
example, \cite{Hartshorne:}). By Serre duality, $H^0(\O(N))$ is dual
to $H^3(\O(-N))=\Ext^3(\O,\O(-N))$. Thus every example of a map
$f$ has some ``dual'' map $f^\vee\in\Ext^3(\O,\O(-N))$.

In terms of the derived category, an element of $\Ext^3(\O,\O(-N))$ is
a morphism 
\begin{equation}
\xymatrix@1{
\underline{\O}\ar[r]^-{f^\vee}&\underline{\O(-N)}[3],
}
\end{equation}
where, recalling the notation of section~\ref{ss:top},
$\underline{\O(-N)}[3]$ refers to a complex whose only nontrivial
entry is $\O(-N)$ at position $-3$.
The cone of this map $\cX_N = \Cone(f^\vee:\underline{\O}\to
\underline{\O(-N)}[3])$ is an object in $\DC(X)$ we wish to study in
this section. Note that we may deform $\cX_N$ into a whole family of
D-branes by varying $f^\vee$ --- just as 4-branes may be deformed by
varying $f$.

\iffigs
\begin{figure}[t]
  \centerline{\epsfxsize=15cm\epsfbox{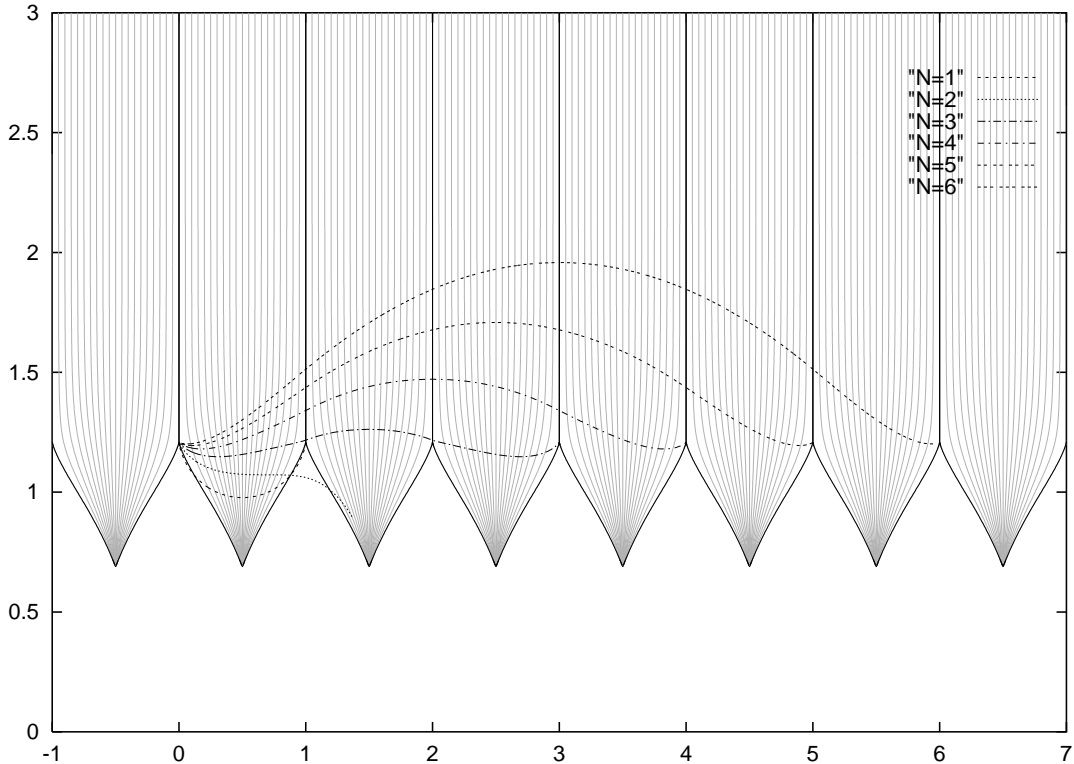}}
  \caption{Stability of the exotic objects $\cX_N$ in the $t$-plane.}
  \label{f:XN}
\end{figure}
\fi

The rules of section~\ref{s:pistab} require that $\grade(\cE[n])=\grade(\cE)
+n$. It follows that the condition for stability of $\cX_N$ is that
\begin{equation}
  3+\grade(O(-N))-\grade(O) < 1.
\end{equation}
This condition can be analyzed in the same way as the previous section
and the result is shown in figure~\ref{f:XN}. This time the region of
stability lies {\em below\/} the marginal line. That is, these exotic
objects only exist at small radius.
Clearly something odd has happened for $N=2$ which is the subject of
the next section.

As before, for large $N$ one may use (\ref{eq:LRZ}) directly to show
that asymptotically the line of stability becomes an arc of a circle
of radius $N/\sqrt{3}$ whose centre this time is located at $t=N/2 -
iN/2\sqrt{3}$.

It is perhaps worth emphasizing that our discussion of the exotic
objects $\cX_N$ requires $\grade$ not be to confined to an interval of
width 2. The stability of the $\cX_N$'s below the line of marginal
stability occurs precisely because the relative grades of
the sheaves $\O$ and $\O(-N)$ exceeds 2! 


\subsection{The peculiar case of $\cX_2$}  \label{ss:X2}

\iffigs
\begin{figure}[t]
  \centerline{\epsfxsize=15cm\epsfbox{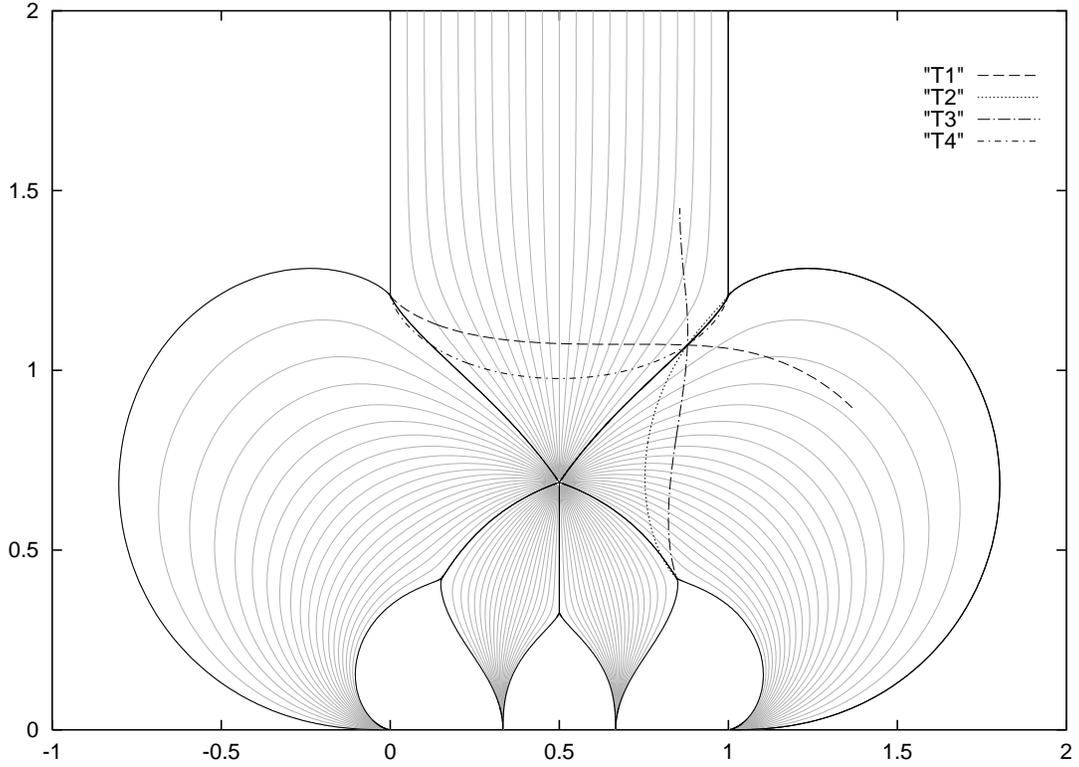}}
  \caption{Lines of stability for the case of $\cX_2$.}
  \label{f:X2}
\end{figure}
\fi

The first thing to notice about the $N=2$ line in figure~\ref{f:XN} is
that we have used an inappropriate set of fundamental regions. The
line of marginal stability passes outside the
fundamental region $0<B<1$ {\em below\/} the conifold point. It is
more appropriate the use Candelas et al's ``scorpion'' \cite{CDGP:} as
shown in figure~\ref{f:X2}. This consists of the 5 fundamental domains
around a Gepner point at $t=\ff12+\ff i2\cot\ff{\pi}5$.

The marginal line for $N=2$ is shown as ``T1'' in figure~\ref{f:X2}
for reasons to become clear shortly. The next striking feature of this
line is that it appears to end to the right at a point of no obvious
significance. This point is where the period of $\O(-2)$ vanishes but
it does not correspond to a conifold point or any other point where
the conformal field theory may diverge. If a physical D-brane becomes
massless, the associated conformal field theory must diverge. It must
therefore be that the D-brane associated to $\O(-2)$ (and hence
$\underline{\O(-2)}[3]$) becomes unstable as we move rightwards
along the line.

We are now in the situation discussed in section \ref{ss:stab}. In
analyzing the stability of $\cX_2$ we need to concern ourselves with
the stability of one of its decay products
$\underline{\O(-2)}[3]$. One possible decay mode of
$\underline{\O(-2)}[3]$ which fits in nicely with this picture is as
follows.

On $\P^4$ the following sequence is exact for $0\leq n\leq5$:
\begin{equation}
0\to\Omega^n(n)\to\O^{\oplus\binom5n}\to
\O(1)^{\oplus\binom5{n-1}}\to\ldots\to\O(n)\to0,
	\label{eq:Os}
\end{equation}
where $\Omega^n$ is the $n$th exterior power of the cotangent sheaf.
Putting $n=3$ and $n=5$ into this formula and comparing results we
obtain the short exact sequence
\begin{equation}
0\to\O(-2)\to\O(-1)^{\oplus5}\to\Omega^3(3)\to0. \label{eq:T2}
\end{equation}
As these sheaves are locally free, this sequence remains exact when we
restrict to the quintic hypersurface $X$.
We implicitly perform this restriction from now on.\footnote{Note in particular
that $\Omega$ refers to the cotangent sheaf of $\P^4$
restricted to $X$ --- not the cotangent sheaf of $X$!}
This short exact sequence gives a distinguished triangle with
$\underline{\O(-2)}[3]$ at one vertex as desired. There is good reason
to believe that both $\O(-1)$ and $\Omega^3(3)$ are stable in the
region we are considering. This is because they are the so-called
``fractional branes'' of \cite{DD:stringy}. 
When these branes become massless they
are responsible for the singularities of the conformal field theories
associated to the corresponding conifold points in figure~\ref{f:X2}.

\begin{figure}
\begin{equation}
\begin{xy}
\xymatrix{
&\O(-2)[3]\ar[lddd]\ar[rd]&\\
\Omega^3(3)[3]\ar@{.>}[dd]_{[1]}\ar[ur]^{[1]}&&\O(-1)^5[3]
        \ar[ll]\ar@{.>}[dddl]\\
&&\\
\cX_2\ar[rr]^{[1]}\ar@{.>}[dr]&&\O\ar[uuul]\ar@{.>}[uu]\\
&\Scr{Y}\ar@{.>}[ur]_{[1]}\ar@{.>}[uuul]&
}
\POS<26mm,-26mm>*{\txt{\scriptsize T1}}
\POS<26mm,-10mm>*{\txt{\scriptsize T2}}
\POS<8mm,-36mm>*{\txt{\scriptsize T3}}
\POS<44mm,-36mm>*{\txt{\scriptsize T4}}
\end{xy}
\end{equation}
\caption{The octahedron for $\cX_2$.} 
\label{f:oct2}
\end{figure}

The relevant octahedron is shown in figure~\ref{f:oct2} (where we omit
various underlines for clarity). The triangle ``T1'' is the original
triangle defining $\cX_2$, and ``T2'' comes from (\ref{eq:T2}). 
``T4'' now defines the D-brane $\Scr{Y}$ and determines its
stability. Finally ``T3'' allows for a decay of $\cX_2$ into $\Scr{Y}$ and
$\underline{\Omega^3(3)}[3]$. The corresponding four lines of
marginal stability are shown in figure~\ref{f:X2}. Note that the
analysis of section~\ref{ss:stab} shows in general that the point of
intersection of T1 and T2 will also lie on T3 and T4.

Now we have a consistent picture for the stability of $\cX_2$. It is
stable (at least with respect to the triangles we have considered)
{\em below\/} the T1 line and to the {\em left\/} of the T3 line. The
rightmost point of the T1 line, and similarly the top of the T3 line,
have no physical significance as $\cX_2$ decays before they can be reached.


\section{Monodromy}  \label{s:mon}

\subsection{A point on the quintic}  \label{ss:ptq}

We have seen above that the properties of a D-brane are not purely a
function of the moduli space but rather the universal cover of the
moduli space, or the Teichm\"uller space. The
case of $\O(-2)$ clearly illustrates this. In section~\ref{ss:D4} we
saw how $\O(-2)$ becomes massless at the conifold point where
$B=2$. In section~\ref{ss:X2} we saw it instead becomes massless at a
point in the interior of the moduli space.

Of course, two fundamental regions of the moduli space must represent
identical physics. The set of D-branes and their physical properties is
determined completely by a point in the moduli space. It can only be a
given D-brane's interpretation as an element of $\DC(X)$ that depends
on how we got there, or equivalently, by a point in the Teichm\"uller
space.

This leads immediately to the conclusion that monodromies in the
moduli space induce {\em autoequivalences\/} of
$\DC(X)$. Such an autoequivalence is always induced by a Fourier--Mukai
transform \cite{Orl:FM}. Fixing notation, given projections
\begin{equation}
\xymatrix{
  &X\times X\ar[dl]_{p_1}\ar[dr]^{p_2}\\
  X&&X
}
\end{equation}
an object $\Psi\in\DC(X\times X)$ induces a transform 
\begin{equation}
  T_{\Psi}(A)=\mathbf{R}p_{2*}(\Psi\Lotimes p_1^*A),  \label{eq:FM}
\end{equation}
for any object $A$ of $\DC(X)$.

Based on the action on the Chern characters,
Kontsevich \cite{Kont:mon} conjectured that for the monodromy around the
{\em conifold\/} point in the case of the quintic, $\Psi$ is given by
\begin{equation}
\ldots\to0\to\O_{X\times X} \to \underline{\O_\Delta}\to 0\to\ldots,
     \label{eq:Psi0} 
\end{equation}
where $\Delta\subset X\times X$ is the diagonal embedding of $X$, and
the map in (\ref{eq:Psi0}) is the obvious restriction map. Seidel and
Thomas \cite{ST:braid} analyzed many aspects of this transform in the
context of ``homological mirror symmetry''.  Indeed they outlined an
interesting framework in this context in which one can motivate
Kontsevich's conjecture.  (Note that we will not directly appeal to
this conjectured strong form of mirror symmetry in this paper.)

Note that (\ref{eq:FM}) can be written in a simpler form involving only
$\DC(X)$: 
\begin{equation}
T_\Psi(A) = \Cone(e:\Hom(\underline{\O},A)\otimes\underline{\O}\to A),
	\label{eq:ST}
\end{equation}
where $e$ is the evaluation map. We refer to \cite{ST:braid} for the
precise definitions of the manipulations in (\ref{eq:ST}). 

We are now in a position to justify Kontsevich's proposal at the level
of the derived category, rather than just the cohomology class of
Chern characters. We begin with the 0-brane $A=\underline{\O_p}$,
i.e., the skyscraper sheaf at a point $p\in X$ at zeroth position in
the complex.

\iffigs
\begin{figure}[t]
  \centerline{\epsfxsize=13cm\epsfbox{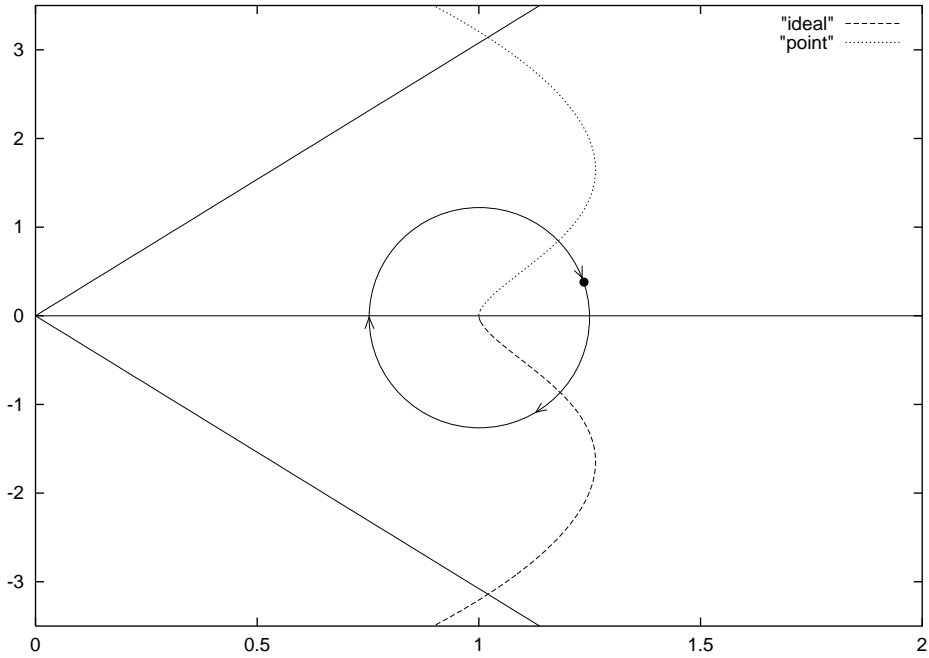}}
  \caption{Lines of stability for skyscrapers and ideals in the $\psi$-plane.}
  \label{f:idpt}
\end{figure}
\fi

Let $\cI_p$ denote the ideal sheaf of functions vanishing at $p$.
The short exact sequence
\begin{equation} \label{eq:idealseq}
 0\to\cI_p\to\O\to\O_p\to0,
\end{equation}
induces the distinguished triangle
\begin{equation}
\xymatrix{
&\underline{\cI_p}[1]\ar[dl]^w_{[1]}&\\
\underline{\O}\ar[rr]^u&&\underline{\O_p}.\ar[ul]^v
} \label{eq:idpt}
\end{equation}
The grades of these objects can be calculated as before given the
Chern characters
\begin{equation}
\begin{split}
\ch(\O) &= 1\\
\ch(\O_p) &= \ff15e_1^3\\
\ch(\cI_p) &= 1-\ff15e_1^3.
\end{split}
\end{equation}

Begin at a point near large radius limit and assume $-1<B<0$, i.e., 
$0<\arg\psi<2\pi/5$. Now loop clockwise around the conifold point at
$\psi=1$ as shown in figure~\ref{f:idpt}. The straight lines in this
figure denote the boundaries of fundamental regions of the moduli
space along $\arg\psi=2\pi n/5$. The same path is
shown in figure~\ref{f:idptBJ} in the $t$-plane.

\iffigs
\begin{figure}[t]
  \centerline{\epsfxsize=13cm\epsfbox{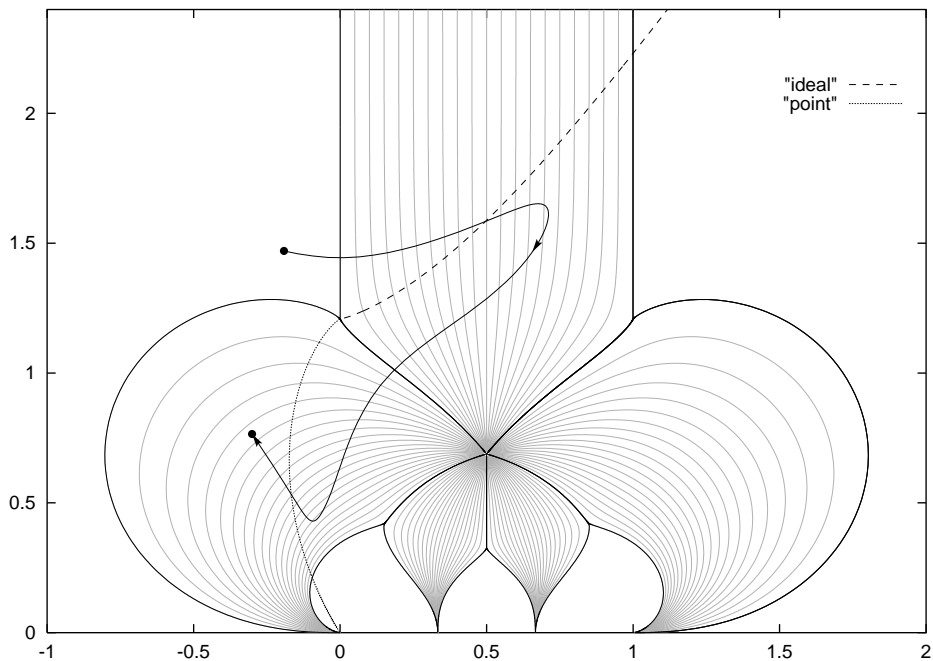}}
  \caption{Lines of stability for skyscrapers and ideals in the $t$-plane.}
  \label{f:idptBJ}
\end{figure}
\fi

At large radius limit, when we begin our journey, it is easy to show
that $\grade(\O_p)-\grade(\O)=\ff32$. Therefore the string on side $u$ of
(\ref{eq:idpt}) is massive and so $\underline{\cI_p}[1]$ is unstable.
This is a quick way of showing the instability of the D0-D6 system as
discussed in \cite{Lif:loop,Shein:06}.

The first interesting thing that happens as we follow the path is that 
the string on side $u$ becomes tachyonic thus stabilizing
$\underline{\cI_p}[1]$. This line of marginal stability is labeled
``ideal'' in figures \ref{f:idpt} and \ref{f:idptBJ}. The ideal sheaf
is stable to the {\em left\/} of this line in figure \ref{f:idpt}.

Continuing along the path the next interesting event occurs when the
string along side $w$ of (\ref{eq:idpt}) becomes massive thus
destabilizing the 0-brane $\underline{\O_p}$. This is exactly the
instability in the 0-brane anticipated in \cite{AL:DC}. The line of
marginal stability is shown as ``point'' in figures \ref{f:idpt} and
\ref{f:idptBJ}. The 0-brane is stable to the {\em left\/} of this
line.

The net result is that our set of stable D-branes has gained
$\underline{\cI_p}[1]$ and lost $\underline{\O_p}$. It is not hard to
see that this course of events is entire due to the fact that the
period associated to $\O$ had a simple zero at the conifold
point. That is, it is intrinsically associated to monodromy around the
conifold point.

It is therefore natural to assert that monodromy around the conifold
point should be associated to {\em replacing\/} the 0-brane
$\underline{\O_p}$ by the ideal sheaf $\underline{\cI_p}[1]$. This is
exactly the result predicted by~(\ref{eq:ST}) since
\begin{equation}
\begin{split}
  T_\Psi(\underline{\O_p}) &= \left( \O\to\underline{\O_p} \right)\\
   &= \underline{\cI_p}[1].
\end{split}
\end{equation}

Note that this example provides the type of phenomenon promised in
section \ref{ss:bun}. At one point in the moduli space the object 
$\underline{\cI_p}[1]$ decays into $\underline{\O_p}$ and $\underline{\O}$
whereas elsewhere in the moduli space $\underline{\O_p}$ decays into
$\underline{\cI_p}[1]$ and $\underline{\O}$. This shows the
``nonabelianness'' of the category of D-branes.

This example displays an important difference from the behavior of the
BPS spectrum in pure $SU(2)$ SYM (and many other examples), as
described in \cite{SW:I,Fzdn:msline,AFS:msline}.  The analogy is of course that
both problems exhibit ms-lines passing through a point at which a
single BPS state goes massless, on which generic objects with the
conjugate charge decay.  The difference is that in SYM, this part of
the spectrum is smaller in the strong coupling regime; the $W$ boson
and generic dyons decay in this regime, leaving a single stable
massive conjugate brane (with its antibrane).  In the present example,
the spectrum is larger in the stringy (strongly coupled sigma model)
regime.  In particular, on the left of the ms-lines in figure
\ref{f:idpt}, {\it both} the original D$0$ $\O_p$ {\it and} its
monodromy image $\cI_p[1]$ are stable.  This is a simple consequence
of the different behaviors of the central charges in the two problems,
but shows that overly simple generalizations about the qualitative
behavior of the spectrum can fail.

This example is also interesting from the point of view of
``D-geometry'' as discussed in \cite{Doug:Dgeom}.  One of the
prototypical problems one can study to get a handle on stringy
geometry is the moduli space metric of the D$0$-brane; this
D$0$-metric is a canonically defined metric which for \CY\ exhibits
the well-known corrections to Ricci-flatness \cite{GVZ:4loop},
assuming that the D$0$ exists in the stringy regime.  What we have
found is that the D$0$ not only (almost certainly) exists,\footnote{
Although we did not consider decays into states with D$2$ or D$4$
charge, this is implausible for a variety of reasons, the simplest
being that such states which are known to exist have larger mass than
the D$0$ near the conifold point. The stability of the 0-brane has
also been anaylzed in \cite{Schd:D0}.} but that in the stringy regime
there are two candidate D$0$'s, which can be obtained by going around
either side of the conifold point, leading to two D$0$-metrics.  It
would be interesting to know if these are necessarily the same or are
related in some simple way.


\subsection{The general case}  \label{ss:mgen}

Given the above example we can try to generalize to the case of any
object in $\DC(X)$ for the quintic threefold $X$. We know that the
monodromy must induce an autoequivalence on $\DC(X)$ and must
therefore arise as a Fourier--Mukai transform associated to some
$\Psi\in\DC(X\times X)$. 

Putting $A=\underline{\O_p}$ in~(\ref{eq:FM}) yields
\begin{equation}
\begin{split}
  T_\Psi(\underline{\O_p}) &= \mathbf{R}p_{2*}\left(\Psi|_{p\times
                                X}\right)\\
	&= \Psi|_{p\times X},
\end{split}
\end{equation}
where $|_{p\times X}$ is the derived category version of the
restriction map (i.e.,
$\displaystyle{\Lotimes(\underline{\O_p}\boxtimes\underline{\O}})$). 
We therefore have that
\begin{equation}
  \Psi|_{p\times X} = \left( \O\to\underline{\O_p} \right).
	\label{eq:Psicon}
\end{equation}
The most natural and obvious solution to this equation for $\Psi$ is that
\begin{equation} \label{eq:kontsev}
  \Psi = \left( \O_{X\times X}\to\underline{\O_\Delta} \right)
\end{equation}
as predicted by Kontsevich. We will establish this as the only
possibility shortly.

The argument we gave in section~\ref{ss:ptq} generalizes directly to any
object $E$ satisfying
\begin{equation} \label{eq:onehom}
1 = \sum_i \dim \Hom(\O,E[i]) .
\end{equation}
Given that there is a unique morphism $f$ between $\O$ and $E$ in some
degree, there is a unique triangle involving these two objects and $\Cone(f)$
which will play the role of (\ref{eq:idealseq}).  Following the loop around
the conifold point will again stabilize $\Cone(f)$ and destabilize $E$,
again motivating the claim that (\ref{eq:ST}) is the monodromy image of $E$.

Note that $\O$ itself is not actually invariant upon moving around the
conifold point. One can show that
\begin{equation}
  T_\Psi(\underline{\O}) = \underline{\O}[-2] ,
\end{equation}
consistent with the claim that the combination of this FMT and flow
of gradings is a physical invariance.

The same argument predicts that any object $E$ with non-zero
$\Hom(\O,E[i])$, not necessarily satisfying (\ref{eq:onehom}), will
decay, but now one does not get a unique candidate to replace it.
Besides the evaluation map $e$ used in (\ref{eq:ST}), one will in
general form other objects as bound states of $E$ and $\O$ which are
cones of other (non-canonical) maps, such as the individual elements
of $\Hom(\O,E[i])$ or other direct sums of these.  While adiabatically
varying the moduli around the conifold point should lead to a unique
decay process $E \to E' + n \O$, it is not completely clear to us
that $E'$ must be the monodromy image in the sense we are discussing now; it
could be that such an adiabatic process in fact changes the physical
identity of the brane.

However, by combining physical and mathematical input, one can
still derive Kontsevich's conjecture.  Specifically, we need
(\ref{eq:Psicon}), the action of the monodromy on the Chern character,
and the statement that the monodromy is an autoequivalence.
We will appeal to a result of Bridgeland and Maciocia \cite{BM:FMq} (section
3.3), which states the following. Suppose $\Phi:\DC(Y)\to\DC(X)$ is a FMT 
such that for each $y\in Y$, there is a point $f(y)\in X$ with
\begin{equation}
\Phi(\O_y) = \O_{f(y)}.
\end{equation}
This induces a morphism $f:Y\to X$ and furthermore the FMT is
necessarily of the form
\begin{equation}
\Phi(E) = f_*(\underline{L} \Lotimes E)
\end{equation}
for any $E\in\DC(Y)$,
where $L$ is a line bundle on $Y$, and $f_*$ is the direct image functor.

By composing the true monodromy associated to the conifold point with
the inverse of the FMT (\ref{eq:ST}), one obtains an autoequivalence which,
under our assumptions, is an FMT that preserves the Chern
character and every point sheaf $\O_p$.
This FMT is covered by the result we just cited with $f(y)=y$.
Using the action on the charges to
determine $L=\O_Y$, we find that this composition is indeed the identity,
establishing the result.

Let us turn to some related comments.
It is also instructive to follow $\O$ as we go  around the Gepner
point five times. This and its generalizations have been analyzed at
the Chern character level in papers such as
\cite{Mor:geom2,Horj:DX,Tomas:McK,GJ:McK,Mayr:McK,me:navi} but not at
the derived category level.

Monodromy around the large radius limit is trivially achieved by
tensoring with $\O(1)$. One may then consider the combined monodromy
transform given by
\begin{equation}
  \Psi_G = \left( \O(1)\boxtimes\O\to\underline{\O_\Delta(1)} \right)
\end{equation}
for the combined monodromy around the large radius limit and conifold
point. That is, $\Psi_G$ yields monodromy around the Gepner point. One
can then compute
\begin{equation}
\begin{split}
  T_{\Psi_G}(\underline{\O}) &= \left(\O^{\oplus5}\to\underline{\O(1)}\right)\\
  (T_{\Psi_G})^2(\underline{\O}) &= \left(\O^{\oplus10}\to
	\O(1)^{\oplus5}\to\underline{\O(2)}\right)\\
  (T_{\Psi_G})^3(\underline{\O}) &= \left(\O^{\oplus10}\to\O(1)^{\oplus10}\to
	\O(2)^{\oplus5}\to\underline{\O(3)}\right)\\
  (T_{\Psi_G})^4(\underline{\O}) &= \left(\O^{\oplus5}\to
	\O(1)^{\oplus10}\to\O(2)^{\oplus10}\to
	\O(3)^{\oplus5}\to\underline{\O(4)}\right).
\end{split}
\end{equation}
The restriction of (\ref{eq:Os}) to $X$ can then be used to show that
$(T_{\Psi_G})^4(\underline{\O}) = \underline{\O(-1)}[4]$. A final
application of $T_{\Psi_G}$ then yields the result
\begin{equation}
  (T_{\Psi_G})^5(\underline{\O}) = \underline{\O}[2].
\end{equation}
Indeed it is an unpublished result \cite{Gepmon:} that 
\begin{equation}
  (T_{\Psi_G})^5(A) = A[2],
\end{equation}
for any $A\in\DC(X)$. In particular ``5 times around the Gepner
point'' is not the identity map on $\DC(X)$ but rather a shift. A
similar result for the elliptic curve was described in
\cite{ST:braid}.

The analysis of the quintic should extend naturally to any \CY\
manifold represented 
as a complete intersection in a toric variety. It particular, there is
a ``primary component of the discriminant locus'' in the moduli space
which plays the r\^ole of the conifold point above as discussed by Horja
\cite{Horj:DX} and Morrison \cite{Mor:geom2} and also in
\cite{me:navi}. Monodromy around this will then also be given by 
the object in (\ref{eq:Psi0}).

What about other monodromies for a general \CY\ threefold? 
The monodromy around the conifold point arises purely because of the
simple zero appearing in $Z(\underline{\O})$. It is clear then
that the arguments above should generalize easily for any case where a
single soliton is becoming massless at a given point in the moduli
space. This is characteristic of all ``conifold''-like points as argued in
\cite{Str:con}. We simply replace $\underline{\O}$ by the particular
object $S\in\DC(X)$ becoming massless. This yields
\begin{equation}
  \Psi = \Cone\left(S^\vee\mathrel{\mathop\boxtimes^{\mathbf{L}}}
	S\to\underline{\O_\Delta}\right),
\end{equation}
as the corresponding object generating the Fourier--Mukai
transform. It would be nice to establish the more general monodromies
discussed by Horja \cite{Horj:EZ}.

\section{An Application to Supersymmetry Breaking}  \label{s:susy}

Let us conclude by briefly describing one possible application of
these ideas, namely to finding $\CN=1$ compactifications of string
theory which spontaneously break supersymmetry.

In the general context of compactification with branes, a popular way
to get supersymmetry breaking is to incorporate combinations of BPS
space-filling branes preserving different supersymmetries, for example
a brane and its antibrane.  One could incorporate non-BPS branes for
the same purpose.

Let us discuss this type of construction in the context at hand of
tree level type II compactification.  As is well known, in this
limit BPS brane considerations do not depend on the number of
Minkowski dimensions filled by the branes, so we can apply
$\Pi$-stability as discussed here to this problem, and say that a
compactification will break supersymmetry in this way if the K theory
class of the embedded D-branes do not support a stable or semi-stable
configuration.  By semi-stable we mean a direct sum of BPS branes with
aligned central charge, so this is just a rephrasing of the problem.

The value of this rephrasing is that, while it is not hard to find
combinations of BPS branes which break supersymmetry for a particular
value of the \CY\ and brane moduli, it is quite hard to find
combinations which break supersymmetry for all values of the moduli.
Indeed we are not aware of any example in the literature.  Since the
\CY\ moduli are of course dynamical fields in the compactified theory,
one must remove supersymmetric vacua at all points in moduli space to
truly break supersymmetry.

To give the simplest example of what can go wrong, if one includes a
brane and its own antibrane, typically after varying moduli the two
will annihilate, leaving a supersymmetric configuration.  
This is typically signaled by a tachyon in the
spectrum from the start and is thus not so interesting.  Many examples
are known in which no tachyon is visible in the spectrum, at least for
the particular \CY\ moduli for which the configuration was constructed,
and yet supersymmetry is broken.  We discussed the familiar example of
the D$0$-D$6$ system in section 4, and the techniques here provide
many generalizations of this.

The D0-D6 configuration however does not pass the test of breaking
supersymmetry for all values of the moduli; on the ms-line we
exhibited in figure 7, supersymmetry is restored.  This phenomenon is
very generic: for example, any compactification with two types of BPS
brane would be expected to restore supersymmetry on a wall of real
codimension 1.  This also points to the larger problem, which is that
since BPS central charges vary drastically with the moduli, an
analysis in the neighborhood of some point in moduli space is not at
all adequate to address this problem.  A global picture is needed.

We note in passing that a consistent string compactification with
D-branes must of course include orientifold planes and cancel RR
tadpoles, and that one would be particularly interested in finding a
brane configuration of this type which breaks supersymmetry.  At
present it appears that breaking supersymmetry is the hard part of the
problem, while numerous examples of tadpole canceling configurations
are known.

Using the ideas discussed in subsection 2.6 and in \cite{DFR:orbifold},
one can see that in the $\C^3/\Z_3$ orbifold (and probably in general
$\C^3/\Gamma$ orbifolds), one can never break supersymmetry in this
way.  In other words, for every RR charge vector, there is some point in
Teichm\"uller space $\cT$ at which the configuration is semi-stable,
and supersymmetry is restored.

To analyze this problem, it is simplest to express the RR charges in the
basis of fractional branes discussed in \cite{DG:fracM,DFR:orbifold}.
The simplest case to consider is then when all charges in this basis are
non-negative, or all are non-positive, as then it is manifest that 
supersymmetry is restored at the orbifold point: one simply considers
the appropriate direct sum of fractional branes, all of which
preserve the same supersymmetry.
The result holds even for general charges, because by moving
on the Teichm\"uller space one can induce a monodromy after which
all charges become positive in this basis.  It suffices to consider the
monodromy corresponding to $B\to B+n$ for sufficiently large $n$; by
results cited in \cite{DFR:orbifold} the resulting charge vector is
$n_i = H^0(M,V \otimes O(i-3+n)) \ge 0$ for $i=1,2,3$ and is manifestly
non-negative.  Thus there will exist monodromy images of the ``original''
orbifold point at which supersymmetry is restored.

The main ingredient which is needed to get this very general result is
that there exist a point in moduli space at which a set of branes forming
a basis for the entire derived category of branes (such as the fractional
branes in the orbifold problem) all have aligned central charges.
As we discussed in subsection 2.6, this is probably not true for compact
\CY, leaving the possibility of supersymmetry breaking by this mechanism
open.  It would be quite interesting to find a concrete example.


\section*{Acknowledgments}

It is a pleasure to thank P.~Candelas, S.~Katz, M.~Kontsevich,
A.~Lawrence, C.~McMullen, D.~Morrison, R.~Plesser,
E.~Sharpe, R.~Thomas and M.~Vybornov for useful conversations. 
P.S.A.~is supported in part by NSF grant DMS-0074072 and by a research
fellowship from the Alfred P.~Sloan Foundation. 
M.R.D. is supported by DOE grant DE-FG02-96ER40959.
This collaboration was supported by an NSF FRG grant
(DMS-0074072 and DMS-0074177).


\end{document}